\documentclass[12pt,amsmath]{iopart}
\usepackage{graphicx}
\usepackage{amssymb}

\newcommand{\Q}{\ensuremath{\mathcal{Q}}}

\newcommand{\vc}{\mathbf}
\newcommand{\stres}{\ensuremath{\Pi}}
\newcommand{\ident}{\ensuremath{\mathcal{I}}}

\newcommand{\LB}{\scriptsize{\textrm{LB}}}

\newcommand{\IE}{\scriptsize{\textrm{IE}}}

\newcommand{\erf}{\textrm{erf}}
\newcommand{\flu}{\scriptsize{\textrm{fl}}}
\newcommand{\kin}{\scriptsize{\textrm{kin}}}

\begin{document}

\title[Kinetic Bohm Criterion]{Kinetic Theory of the Presheath and the Bohm Criterion}

\author{S D Baalrud\footnote{Present address: Department of Physics and Astronomy, University of Iowa, Iowa City, IA 52242} and C C Hegna}

\address{Department of Engineering Physics, University of Wisconsin-Madison, 1500 Engineering Drive, Madison, WI 53706, USA}
\ead{scott-baalrud@uiowa.edu}

\begin{abstract}

A kinetic theory of the Bohm criterion is developed that is based on positive-exponent velocity moments of the plasma kinetic equation. This result is contrasted with the conventional kinetic Bohm criterion that is based on a $v^{-1}$ moment of the Vlasov equation. The salient difference between the two results is that low velocity particles dominate in the conventional theory, but are essentially unimportant in the new theory. It is shown that the derivation of the conventional kinetic Bohm criterion is flawed. Low velocity particles can cause unphysical divergences in the conventional theory. These divergent contributions are avoided with this new approach. The two theories are compared using example distribution functions from previous presheath models.  The importance of ion-ion and electron-electron collisions to determining the particle distribution functions throughout the presheath is also discussed.  A kinetic equation that accounts for wave-particle scattering by convective instabilities is used to show that ion-acoustic instabilities in the presheath of low temperature plasmas (where $T_e \gg T_i$) can cause both ions and electrons to obtain Maxwellian distribution functions near the sheath.

\end{abstract}
\submitto{\PSST}
\maketitle

\section{Introduction\label{sec:intro}}

The Bohm criterion, as originally derived~\cite{bohm:49}, provides a condition that the ion flow speed must satisfy as ions leave a quasineutral plasma and enter a nonneutral sheath. Bohm's derivation assumed a single species of monoenergetic ions with velocity $\vc{V}_i$ and Boltzmann electrons with temperature $T_e$.  From these assumptions, he showed that ions must be supersonic as they leave the plasma: $V_i \geq c_s \equiv \sqrt{T_e / M_i}$~\cite{bohm:49}.  It has since been shown theoretically~\cite{riem:97} and experimentally~\cite{oksu:02} that equality typically holds in the Bohm criterion. Although the assumptions that Bohm made in his seminal work often provide an accurate approximation of physical systems, an important question to answer is: How does the Bohm criterion change for more general electron and ion distribution functions?

Attempts to answer this question have been the topic of several theory papers over the past fifty years \cite{harr:59,hall:62,caru:62,andr:71,alle:76,emme:80,riem:81,biss:87a,biss:87,biss:89,koch:89,proc:90,riem:91,riem:95,amem:97,riem:03,ster:04,fern:05,riem:06,kuhn:06,jeli:07,alle:09}. Harrison and Thompson put forth one of the first such generalizations of the Bohm criterion~\cite{harr:59}. Like Bohm, they assumed that the electron density obeys a Boltzmann relation, but allowed for a general ion distribution, $f_i(\vc{v})$. They predicted that ions should obey the criterion $\int d^3v\, v_z^{-2} f_i (\vc{v})/n_i \leq c_s^{-2}$, at the sheath edge.  Later work by Riemann \cite{riem:81,riem:91} generalized the electron term as well and found the criterion
\begin{equation}
\frac{1}{M_i} \int d^3v\, \frac{f_i (\vc{v})}{v_z^2} \leq - \frac{1}{m_e} \int d^3v\, \frac{1}{v_z} \frac{\partial f_e (\vc{v})}{\partial v_z} . \label{eq:kbohm}
\end{equation}
Equation~(\ref{eq:kbohm}) is commonly called the ``generalized Bohm criterion'' or the ``kinetic Bohm criterion.'' It is a frequently cited result that has been applied in both analytic and numerical models of laboratory~\cite{biss:87a,biss:87,biss:89,koch:89,proc:90,riem:91,riem:95,amem:97,riem:03,ster:04,fern:05,riem:06,kuhn:06,jeli:07,alle:09,jeli:09,ahed:10} and fusion~\cite{stan:84,cohe:04} plasmas. It is also discussed in a popular textbook~\cite{lieb:05}. In deriving equation~(\ref{eq:kbohm}), the electric field of the sheath is taken to be aligned in the $\hat{z}$ direction, and it is assumed that the only spatial gradients of the particle distribution functions are caused by this electric field. 

Although it is frequently cited in the theoretical literature, equation~(\ref{eq:kbohm}) does not produce a meaningful criterion for many plasmas of interest. If the ion distribution function has any particles with zero velocity, the left side of equation~(\ref{eq:kbohm}) diverges. If the velocity gradient of the electron distribution does not vanish at $v_z=0$, the right side of equation (\ref{eq:kbohm}) diverges. Consider the common example where the ion distribution function is a flow-shifted Maxwellian and the electron distribution is a stationary Maxwellian. In this case, the left side of equation (\ref{eq:kbohm}) is infinite and the right side is $n_e/T_e$, which produces an unphysical prediction. Equation (\ref{eq:kbohm}) places undue importance on the part of the distribution functions where particles are slow. 

These shortcomings of equation (\ref{eq:kbohm}) have been pointed out before. Shortly after the publication of Harrison and Thompson's result, Hall discussed the fact that unphysical divergences arise when the ion distribution contains slow particles \cite{hall:62}. Lieberman and Lichtenberg \cite{lieb:05} mentioned that equation~(\ref{eq:kbohm}) can lead to ``mathematical difficulties'' at low energies. However, despite the fact that it often gives unphysical results and that this shortcoming has been discussed in previous publications, equation~(\ref{eq:kbohm}) continues to be used in plasma physics literature. 

Experimental and numerical studies of presheath physics have been published that appear to contradict equation~(\ref{eq:kbohm}). Laser induced fluorescence has been used to measure the ion distribution function in the presheath of laboratory plasmas \cite{bach:95,oksu:01,clai:06}. These experiments found a Maxwellian ion distribution with a flow speed approaching the sound speed at the sheath edge; thus, $f_i(v_z=0) \neq0$. Numerical simulations by Sheridan~\cite{sher:01}, Meige \etal~\cite{meig:07} and Robertson~\cite{robe:09} also find ion distribution functions with populations of low energy ions throughout the presheath. These used hybrid (fluid electron, kinetic ion) and fully kinetic particle-in-cell techniques. Consistent with the experiments, these simulations find that the ion flow speed approaches the sound speed at the sheath edge. These studies apparently contradict the implication of equation~(\ref{eq:kbohm}) that the low energy part of the ion distribution function dominates the ion dynamics at the sheath edge. 

In section~\ref{sec:previous}, we reconsider previous derivations of equation~(\ref{eq:kbohm}). We show that these derivations contain two errors. The first of these is taking the $v_z^{-1}$ moment of the collisionless kinetic equation (i.e. Vlasov equation). Neglecting the collision operator is a mistake because the $v_z^{-1}$ moment of it diverges if the distribution functions contain particles near zero velocity. Not all distribution functions of interest satisfy the Vlasov equation. Only velocity moments with a positive power should be applied, otherwise divergences can arise in the integrand at $v_z=0$ for non-Vlasov solutions. The second error is a mathematical mistake where integration by parts is misapplied to a function that is not continuously differentiable. This error can easily be corrected, but the resultant criterion then differs from equation~(\ref{eq:kbohm}).  

In section~\ref{sec:newbohm}, we derive an alternative kinetic Bohm criterion that is based upon positive exponent velocity moments of the kinetic equation, rather than the $v_z^{-1}$ moment of previous work. This approach avoids the possibility of divergent velocity-space integrals. Particles with low energy do not have any special significance in this theory. It also supports previous derivations of the Bohm criterion based on fluid equations, and it returns these results in the fluid limit. In contrast, equation~(\ref{eq:kbohm}) cannot return fluid results because it places undue importance on low energy particles.  In section~\ref{sec:examples}, we consider some example distribution functions that are common in low temperature plasmas, but for which the generalized Bohm criterion we derive in section~\ref{sec:newbohm} can give significantly different predictions than equation~(\ref{eq:kbohm}). We consider example presheath models with a cold ionization source (the Tonks-Langmuir problem), a warm ionization source as well as a collisional example where ions and electrons are Maxwellian. In section \ref{sec:collisions}, we discuss how electron-electron and ion-ion collisions in the presheath are often important in order to determine the distribution functions near the boundaries of laboratory plasmas. Specifically, we show how ion-acoustic instabilities can significantly enhance these collisions and drive the distribution functions toward Maxwellians. 

\section{The conventional kinetic Bohm criterion\label{sec:previous}}

\subsection{The sheath edge and sheath criterion}

To derive equation~(\ref{eq:kbohm}), one must first develop a mathematical definition of where the quasineutral plasma (presheath) ends and the nonneutral sheath begins. If the Debye length is shorter than the ion momentum transfer collision length ($\lambda_{D} \ll \lambda_{im}$), the presheath and sheath are separated by a transition region~\cite{riem:91}. This transition region scales as $\lambda_{im}^{1/5} \lambda_{D}^{4/5}$~\cite{riem:97,oksu:02}, which is typically much shorter than the presheath length scale ($l$). In the asymptotic limit $\lambda_{D}/l \rightarrow 0$, the transition region is so short that the boundary layer problem can effectively be split into the two regions of quasineutral plasma (presheath and bulk plasma) and nonneutral sheath \cite{riem:91,riem:95,riem:03}. The sharp boundary that results in this limit is what is typically referred to as the ``sheath edge.''

The sheath criterion is a condition that the spatial gradient of the charge density must satisfy at the sheath edge. Riemann derives the sheath criterion by applying to Poisson's equation the physical requirement that quasineutrality breaks down at the sheath edge~\cite{riem:95}.  His derivation can be summarized as follows. Expanding Poisson's equation about the sheath edge, where the electrostatic potential is referenced to zero $\phi = 0$, yields $\nabla^2 \phi = - 4 \pi [ \rho (\phi = 0) + d \rho / d \phi |_{\phi = 0} \phi + \ldots ]$. Here $\rho \equiv \sum_s q_s n_s$ is the charge density. It is assumed that the electrostatic potential variation of the presheath and sheath is one-dimensional and we align this in the $\hat{z}$ direction. At the sheath edge, which is the boundary of quasineutral plasma, the first nonvanishing term in this expansion is the linear term: $d^2 \phi / dz^2 = -4\pi\, d\rho / d \phi |_{\phi=0} \phi$.  Multiplying this by $d \phi / d z$ and integrating with respect to $z$ yields
\begin{equation}
\frac{E^2}{4\pi} + \frac{d \rho}{d \phi} \biggl|_{\phi = 0} \phi^2 = C 
\end{equation}
in which $C$ is a constant. Since $\phi \rightarrow 0 $ as $z/ \lambda_{D} \rightarrow \infty $ on the sheath length scale, the constant $C$ must be zero \cite{riem:95}. We are then left with $d \rho / d \phi |_{\phi = 0} = - E^2 / (4\pi \phi^2)$, which implies the sheath criterion 
\begin{equation}
\frac{d \rho}{d \phi} \biggl|_{\phi = 0} \leq 0. \label{eq:sheathcrit1}
\end{equation}
The sheath criterion provides a mathematical definition of the sheath edge. Using the fact that $d n_s / d \phi  = - E^{-1} d n_s / dz$, the sheath criterion can also be written $\sum_s q_s dn_s / dz |_{z=0} \geq 0$.  Since the relation between density and the distribution function is simply $n_s = \int d^3v\, f_s$, equation~(\ref{eq:sheathcrit1}) becomes
\begin{equation}
\sum_s q_s \int_{-\infty}^\infty d^3v\, \frac{\partial f_s}{\partial z} \geq 0 .  \label{eq:sheathcrit}
\end{equation}

\subsection{Derivation of the conventional kinetic Bohm criterion\label{sec:ckbc}} 

Equation~(\ref{eq:kbohm}) has been derived using various approaches by different authors \cite{alle:76,riem:95}, but each approach contains the following elements. These are collisionless theories based on the one-dimensional steady-state Vlasov equation 
\begin{equation}
v_z \frac{\partial f_s}{\partial z} + \frac{q_s}{m_s} E\, \frac{\partial f_s}{\partial v_z} = 0   \label{eq:1dvlasov}
\end{equation}
where it is assumed that the only spatial gradients of $f_s$ are due to the electric field in the presheath and sheath.  Each approach (at some point in the derivation) also divides equation~(\ref{eq:1dvlasov}) by $v_z$, to obtain an expression for $\partial f_s / \partial z$, which can be put into the sheath criterion of equation~(\ref{eq:sheathcrit}). In other words, the $v_z^{-1}$ moment of the Vlasov equation is used.  Applying this approach yields the following form of a kinetic Bohm criterion
\begin{equation} 
\sum_s \frac{q_s^2}{m_s} \int_{-\infty}^\infty d^3v\, \frac{1}{v_z} \frac{\partial f_s}{\partial v_z} \leq 0  . \label{eq:kbohmspec}
\end{equation} 
Assuming that the plasma consists of a single species of ions with charge $q_i = e$ and electrons, this is 
\begin{equation}
\frac{1}{M_i} \int_{-\infty}^\infty d^3v \frac{1}{v_z} \frac{\partial f_i}{\partial v_z} \leq - \frac{1}{m_e} \int_{-\infty}^\infty d^3v \frac{1}{v_z} \frac{\partial f_e}{\partial v_z}  . \label{eq:kbohmint}
\end{equation}

One additional step is required in order to write equation~(\ref{eq:kbohmint}) in the form of equation~(\ref{eq:kbohm}). This is to integrate the $v_z$ component of the ion term by parts, 
\begin{equation} 
\int_{-\infty}^\infty dv_z \frac{1}{v_z} \frac{\partial f_s}{\partial v_z} = \int_{-\infty}^\infty dv_z \frac{\partial}{\partial v_z} \biggl( \frac{f_i}{v_z} \biggr) + \int_{-\infty}^\infty dv_z\, \frac{f_i}{v_z^2} \label{eq:intbyparts}
\end{equation}
where the surface term (the first on the right side) is taken to vanish since $[f_i/v_z]_{v_z = \pm \infty}=0$ \cite{harr:59,alle:76}. Applying equation~(\ref{eq:intbyparts}) to the left side of equation~(\ref{eq:kbohmint}) completes this simple derivation of the conventional kinetic Bohm criterion from equation~(\ref{eq:kbohm}). 

\subsection{Deficiencies of the conventional kinetic Bohm criterion\label{sec:deficiencies}} 

Two mistakes are made in all derivations of the conventional kinetic Bohm criterion, including the one presented in section~\ref{sec:ckbc}. They are:
\begin{enumerate}
\item The collision operator should not be neglected if one is to take the $v_z^{-1}$ moment of the kinetic equation. This is because $\int d^3v\, C(f_s)/ v_z $ diverges unless the collision operator vanishes. However, the collision operator only vanishes when the plasma is in equilibrium, which implies that both ions and electrons have Maxwellian distributions with the same temperature and flow speed \cite{lena:60,mont:64,baal:08,baal:10}. This would lead to a contradiction in equation~(\ref{eq:kbohm}) because it reduces to $\infty \leq n_e/T_e$ for Maxwellian distributions. 
\item Since the function $(1/v_z) \partial f_s / \partial v_z$ is not generally continuously differentiable, the integration by parts conducted in equation~(\ref{eq:intbyparts}) is invalid. This deficiency has also been pointed out before~\cite{fern:05}.
\end{enumerate}

The easier of these two issues to correct is (ii) since the integration by parts step can simply be avoided and the kinetic Bohm criterion left in the form of equation~(\ref{eq:kbohmspec}) or (\ref{eq:kbohmint}). However, even these equations are incorrect because of issue (i), which will be discussed below. That the integration by parts step of equation~(\ref{eq:intbyparts}) is incorrect can be shown from a simple example. The contentious step is of the form
\begin{equation}
\int_{-\infty}^\infty dx \frac{1}{x} \frac{d f}{dx} = \underbrace{\int_{-\infty}^{\infty} dx \frac{d}{dx} \biggl( \frac{f}{x} \biggr)}_{=0} + \int _{-\infty}^{\infty} dx \frac{f}{x^2}  , \label{eq:ibpex}
\end{equation}
for any physically possible distribution function (e.g., the restrictions $f(\pm \infty) = 0$ and that $f$ is positive for all $x$ can be imposed since any physical distribution must obey these). If one takes as an example $f=\exp(-x^2)$, the left side of equation~(\ref{eq:ibpex}) can be evaluated directly 
\begin{equation}
\int_{-\infty}^\infty dx \frac{1}{x} \frac{d f}{dx} = -2 \int_{-\infty}^\infty dx\, e^{-x^2} = - 2\sqrt{\pi}  .
\end{equation}
However, if the surface term on the right side of equation~(\ref{eq:ibpex}) is taken to be zero, as it is in derivations of the conventional kinetic Bohm criterion, the right side of equation~(\ref{eq:ibpex}) diverges 
\begin{eqnarray}
\int_{-\infty}^{\infty} dx \frac{e^{-x^2}}{x^2} & = & \lim_{\epsilon \rightarrow 0} \biggl( \int_{-\infty}^{-|\epsilon |} dx \frac{e^{-x^2}}{x^2} + \int_{|\epsilon|}^{\infty} dx \frac{e^{-x^2}}{x^2} \biggr) \label{eq:integrationbp} \\ \nonumber
&= & -2 \sqrt{\pi} + \lim_{\epsilon \rightarrow 0} \biggl[ \frac{2}{|\epsilon |} e^{-|\epsilon |^2} + 2 \sqrt{\pi} \textrm{erf} (|\epsilon |) \biggr] \\ \nonumber
& = & -2 \sqrt{\pi} +\lim_{\epsilon \rightarrow 0} \frac{2}{|\epsilon |} e^{-|\epsilon |^2} \rightarrow \infty   .
\end{eqnarray} 
The reason that integration by parts cannot be applied in equation~(\ref{eq:intbyparts}) is that it is only valid for continuously differentiable functions \cite{rudi:64}. However, $v_z^{-1} \partial f/\partial v_z$ is not continuous unless $\partial f/\partial v_z|_{v_z = 0} = 0$, and $v_z^{-1} \partial f/\partial v_z$ is not continuously differentiable unless both $v_z^{-1} \partial^2 f/\partial v_z^2$ and $v_z^{-2} \partial f/\partial v_z$ are continuous. Many physically reasonable, often expected, plasma distribution functions do not satisfy these properties.  Thus, issue (ii) restricts the previous kinetic Bohm criteria to the form of equation~(\ref{eq:kbohmint}). However, issue (i) will show that there is a problem with equation~(\ref{eq:kbohmint}) as well. 

Equation~(\ref{eq:kbohmint}) also places undue importance on the low-velocity part of the distribution functions. The primary deficiency of the collisionless Vlasov approach is simply that the collision operator cannot be neglected if one is interested in $v_z^{-1}$ moments of the kinetic equation.  To illustrate this, consider what happens if the approach of section~\ref{sec:ckbc} is taken but the collision operator is not neglected. Then, the relevant kinetic equation has the form 
\begin{equation}
v_z \frac{\partial f_s}{\partial z} + \frac{q_s}{m_s} E\, \frac{\partial f_s}{\partial v_z} = C(f_s) , \label{eq:pke}
\end{equation}
in which $C(f_s) = \sum_{s^\prime} C(f_s, f_{s^\prime})$ is the total collision operator. The total collision operator consists of the sum of component collision operators describing collisions between the test species $s$ with all species in the plasma ($s^\prime$) including itself ($s=s^\prime$).  The component collision operators have the Landau form~\cite{land:36}
\begin{equation}
C(f_s, f_{s^\prime}) = - \frac{\partial}{\partial \vc{v}} \cdot \int d^3\!v^{\prime}\; \Q \cdot \biggl(\frac{1}{m_{s^\prime}} \frac{\partial}{\partial \vc{v}^\prime} - \frac{1}{m_s} \frac{\partial}{\partial \vc{v}} \biggr) f_s(\vc{v}) f_{s^\prime}(\vc{v}^\prime) , \label{eq:collop}
\end{equation}
in which $\Q$ is a tensor kernel. Lenard \cite{lena:60} and Balescu \cite{bale:60} have shown that in a stable plasma $\Q = \Q_{\LB}$ where
\begin{equation}
\Q_{\LB} = \frac{2 q_s^2 q_{s^\prime}^2}{m_s} \int d^3k \frac{\vc{k} \vc{k}}{k^4} \frac{\delta[ \vc{k} \cdot (\vc{v} - \vc{v}^\prime)]}{\bigr| \hat{\varepsilon}(\vc{k}, \vc{k} \cdot \vc{v}) \bigl|^2}  . \label{eq:qlb}
\end{equation}
Here 
\begin{equation}
\hat{\varepsilon}(\vc{k}, \omega) \equiv 1 + \sum_s \frac{4\pi q_s^2}{k^2 m_s} \int d^3v \frac{\vc{k} \cdot \partial f_s / \partial \vc{v}}{\omega - \vc{k} \cdot \vc{v}} \label{eq:dielec}
\end{equation}
is the plasma dielectric function for electrostatic fluctuations in an unmagnetized plasma. 

Taking the $v_z^{-1}$ moment of equation~(\ref{eq:pke}), in order to find an equation for $\partial f_s / \partial z$, and putting the result into equation~(\ref{eq:sheathcrit}) gives the criterion
\begin{equation}
\sum_s \frac{q_s^2}{m_s} \int_{-\infty}^\infty d^3v \frac{1}{v_z} \frac{\partial f_s}{\partial v_z} \leq \sum_s \frac{q_s}{E} \int_{-\infty}^\infty d^3 v \frac{1}{v_z} C(f_s) , \label{eq:kbohmwithc}
\end{equation}
which can be compared to the Vlasov result from equation~(\ref{eq:kbohmspec}). A brief study of equation~(\ref{eq:kbohmwithc}) shows that not only the left side, but also the right side, which depends on the $v_z^{-1}$ moment of the collision operator, diverges if $\partial f_s / \partial v_z|_{v_z=0} \neq 0$ or $f_s(v_z=0) \neq 0$ for any species $s$. 

Equation~(\ref{eq:kbohmwithc}) shows that neglecting the collision operator is not a consistent approximation when the $v_z^{-1}$ moment is applied. For example, consider a plasma with a stationary Maxwellian electron species and a Maxwellian ion species flowing relative to the electrons. In this case $C(f_i, f_i) =0$ and $C(f_e, f_e)=0$, but $C(f_e, f_i) \neq 0$. Since $f_i(v_z = 0) \neq 0$, the $C(f_e, f_i)$ term will cause the right side of equation~(\ref{eq:kbohmwithc}) to diverge. The ion term on the left side of equation~(\ref{eq:kbohmwithc}) diverges for this example as well. The only case for which the Vlasov approach, and thus equation~(\ref{eq:kbohmint}), is strictly correct is for Maxwellian ions and electrons with the same temperature and flow speeds: In this limit $C(f_e,f_i)=0$. For this case, equation~(\ref{eq:kbohmint}) [or equivalently equation~(\ref{eq:kbohmwithc})] reduces to $n/T - n/T \leq 0$, which is a true statement, but is not useful as a Bohm criterion.  

The presence of an ionization source in the Vlasov equation can produce additional problems for the conventional kinetic Bohm criterion. Including the effects of a source $S_i$ in the above analysis, one finds an equivalent divergent contribution to the right side of equation~(\ref{eq:kbohmwithc}) unless $S_i(v_z=0)=0$. Section~\ref{sec:colesions} discusses the implication for the Bohm criterion of various ionization source functions that have been considered in previous literature. In some of these examples, such as a Maxwellian source function, the resultant ion distribution function contains slow particles causing the conventional kinetic Bohm criterion to diverge. 

\section{A kinetic Bohm criterion from velocity moments of the kinetic equation\label{sec:newbohm}}

By taking positive velocity moments (e.g., $\vc{v}$, $v^2$, $\vc{v} v^2$, $\ldots$) of the full kinetic equation, a formally exact set of fluid equations are derived. However, this approach suffers from the ``closure'' problem where evolution equations for low order moments are determined by higher order moments. Because of this accounting difficulty, the fluid equations are not closed unless further approximations are made. In the limit that the distribution function is a flow shifted Maxwellian, the plasma is completely described by the lowest order moments $(1, \vc{v}$, and $v^2)$, which determine the evolution of $n_s$, $\vc{V}_s$ and $T_s$. However, deviations from Maxwellian can require evaluation of higher order moments in order to track the evolution of each parameter specifying the distribution. 

\subsection{Fluid moments of the kinetic equation\label{sec:fmoms} }

A hierarchy of fluid moment equations can be constructed from velocity moments of the plasma kinetic equation for species $s$
\begin{equation}
\frac{\partial f_s}{\partial t} + \vc{v} \cdot \frac{\partial f_s}{\partial \vc{x}} + \frac{q_s}{m_s} \vc{E} \cdot \frac{\partial f_s}{\partial \vc{v}} = C(f_s) \label{eq:kineticeq}
\end{equation}
by applying the standard definitions of the fluid variables in terms of velocity-space integrals of the distribution function. These are density
\begin{equation}
n_s \equiv \int_{-\infty}^\infty d^3 v\ f_s , \label{eq:density}
\end{equation}
fluid flow velocity
\begin{equation}
\vc{V}_s \equiv \frac{1}{n_s} \int_{-\infty}^\infty d^3 v\ \vc{v} f_s , \label{eq:fluidflow}
\end{equation}
scalar pressure
\begin{equation}
p_s \equiv \int_{-\infty}^\infty d^3 v\ \frac{1}{3} m_s\, v_r^2\, f_s = n_s T_s ,
\end{equation}
stress tensor
\begin{equation}
\stres_s \equiv \int_{-\infty}^\infty d^3 v\ m_s \biggl( \vc{v}_r \vc{v}_r - \frac{1}{3} v_r^2\, \ident \biggr) f_s , \label{eq:stresstens}
\end{equation}
temperature
\begin{equation}
T_s \equiv \frac{1}{n_s} \int_{-\infty}^\infty d^3 v\ \frac{1}{3} m_s v_r^2 f_s = \frac{1}{2} m_s v_{Ts}^2 , \label{eq:temperature}
\end{equation}
and frictional force density
\begin{equation}
\vc{R}_s \equiv \int_{-\infty}^\infty d^3 v\ m_s \vc{v} C(f_s). \label{eq:fric1}
\end{equation}
Here we have defined the relative velocity $\vc{v}_r \equiv \vc{v} - \vc{V}_s$, where $\vc{V}_s$ is the fluid flow velocity from equation~(\ref{eq:fluidflow}). 

The density moment $(\int d^3v \ldots)$ of the kinetic equation~(\ref{eq:kineticeq}) yields the continuity equation
\begin{equation}
\frac{\partial n_s}{\partial t} + \frac{\partial}{\partial \vc{x}} \cdot \bigl( n_s \vc{V}_s \bigr) = 0 . \label{eq:cont}
\end{equation}
The momentum moment $(\int d^3v\, m_s \vc{v} \ldots)$ yields the momentum evolution equation
\begin{equation}
m_s n_s \biggl( \frac{\partial \vc{V}_s}{\partial t} + \vc{V}_s \cdot \frac{\partial \vc{V}_s}{\partial \vc{x}} \biggr) = n_s q_s \vc{E} - \frac{\partial p_s}{\partial \vc{x}} - \frac{\partial}{\partial \vc{x}} \cdot \stres_s + \vc{R}_s . \label{eq:momentum}
\end{equation}
Equations~(\ref{eq:cont}), (\ref{eq:momentum}) and subsequent equations built from higher exponent velocity moments of the kinetic equation constitute a hierarchy of fluid equations.  In the next section, we use equations~(\ref{eq:cont}) and (\ref{eq:momentum}) to formulate a Bohm criterion that can be written in terms of the distribution functions by associating the fluid variables with their definitions in terms of $f_s$ from equations~(\ref{eq:density}) -- (\ref{eq:fric1}). In this way the resultant criterion retains a kinetic form. 

\subsection{A kinetic Bohm criterion\label{sec:bohmcriterion}}

Bohm's original criterion ($V_i \geq c_s$) was a condition concerning the ion speed (assumed to be monoenergetic in his paper) at the sheath edge.  Kinetic Bohm criteria seek to generalize this condition to account for arbitrary ion and electron distribution functions. However, it is unclear exactly what qualifies as a ``kinetic Bohm criterion.''  For instance, the sheath criterion of equation~(\ref{eq:sheathcrit}) specifies a condition that the spatial gradients of the distribution functions must satisfy at the sheath edge, yet it is not typically called a ``Bohm criterion.'' A possible definition for ``Bohm criteria'' might be statements concerning the ion flow speed at the sheath edge. Although it is not obvious that the conventional kinetic Bohm criterion of equation~(\ref{eq:kbohm}) always fits this definition, we will look specifically for a condition concerning the fluid flow velocity of ions [defined in terms of equation~(\ref{eq:fluidflow})] at the sheath edge. 

We assume that the plasma is in steady-state and that the only spatial variation in $f_s$ is due to the electric field in the sheath and presheath. We take this electric field to be in the $\hat{z}$ direction. Thus, the fluid variables $n_s$, $\vc{V}_s$, $T_s$, $p_s$, $\Pi_s$ and $\vc{R}_s$ are only functions of the spatial variable $z$. With these assumptions, the continuity equation~(\ref{eq:cont}) and momentum equation~(\ref{eq:momentum}) reduce to
\begin{equation} 
n_s \frac{d V_{z,s}}{dz} + V_{z,s} \frac{dn_s}{dz} = 0 \label{eq:1dcont}
\end{equation}
and
\begin{equation}
m_s n_s V_{z,s} \frac{d V_{z,s}}{dz} = n_s q_s E - \frac{d p_s}{dz} - \frac{d\, \Pi_{zz,s}}{dz} + R_{z,s} . \label{eq:longmom}
\end{equation}
in which the $z$ subscript refers to the $\hat{z}$ component of a vector and the subscript $zz$ to the $\hat{z}\hat{z}$ component of a tensor. 

Solving equation~(\ref{eq:1dcont}) for $dn_s/dz$ and putting the result into the sheath criterion $\sum_s q_s dn_s/dz |_{z=0} \geq 0$ [equation~(\ref{eq:sheathcrit})] yields
\begin{equation}
\sum_s q_s \frac{n_s}{V_{z,s}} \frac{d V_{z,s}}{dz} \biggl|_{z=0} \leq 0 . \label{eq:dvdzsheath}
\end{equation}
Equation~(\ref{eq:dvdzsheath}) is a condition concerning the spatial gradient of $V_s$ at the sheath edge. We are interested in a condition on $V_s$ itself, which can be obtained by using equation~(\ref{eq:longmom}) to find an expression for $d V_{z,s} / dz$.  Putting equation~(\ref{eq:longmom}) into (\ref{eq:dvdzsheath}) yields 
\begin{equation}
\sum_s q_s \biggl[ \frac{q_s n_s - \bigl( n_s\, dT_s/dz + d \Pi_{zz,s} / dz - R_{z,s} \bigr)/E}{ m_s V_{z,s}^2 - T_s} \biggr]_{z=0} \leq 0 .  \label{eq:longbohmcrit}
\end{equation}

Equation~(\ref{eq:longbohmcrit}) is a kinetic Bohm criterion that provides a condition that the flow speed of ions must satisfy at the sheath edge. It makes no assumptions about the distribution functions and can be written explicitly in terms of them by substituting the fluid variables with their definitions from equations~(\ref{eq:density}) -- (\ref{eq:fric1}). It does depend on spatial gradients of the higher-order moments temperature and stress tensor and the collisional friction. These could be eliminated in terms of spatial gradients of even higher order fluid moments, the heat flux in this case, by using the temperature evolution equation [obtained from the $(\int d^3v\, v^2 \ldots)$ moment of the kinetic equation]. However, no matter how far one carries out the hierarchy expansion, the subsequent Bohm criterion will still depend on a spatial derivative of $f_s$ inside of some fluid moment integral.  Various closure schemes can be applied to deal with these higher-order moments~\cite{kuhn:06}. 

In many plasmas, but not all (see e.g. section~\ref{sec:trun}), the temperature and stress moments vary on spatial scales much longer than the Debye length. For instance, the gradient length scales for the temperature and stress terms are often characteristic of the length scale for collisions between species $s$ and the background neutrals or another species ($s^\prime \neq s$). These lengths are typically on the order of the presheath length scale ($l$). Equation (\ref{eq:longbohmcrit}) can be simplified significantly for such plasmas because the terms in parentheses become negligible.  The friction term is also controlled by collisions and acts over a collision length scale (including friction with neutrals~\cite{fran:03}).  One the other hand, the gradient length scale of the electrostatic potential approaches the Debye length at the sheath edge. Thus, the terms in parentheses in equation~(\ref{eq:longbohmcrit}) are $\mathcal{O}(\lambda_{D}/l) \ll 1$ smaller than the $q_s n_s$ term in such plasmas.  In fact, many presheath models are based on the limit $\lambda_{D}/l \rightarrow 0$, in which case the electric field is found to become infinite at the sheath edge and the terms in parentheses in equation~(\ref{eq:longbohmcrit}) formally vanish \cite{biss:87,riem:91}. If these terms are negligible, equation~(\ref{eq:longbohmcrit}) reduces to
\begin{equation}
\sum_s \frac{ q_s^2\, n_{s}}{m_s V_{z,s}^2 - T_{s}} \biggl|_{z=0}  \leq 0   . \label{eq:newkbohm}
\end{equation}

Equation~(\ref{eq:newkbohm}) can be further simplified if the electron fluid flow speed is much slower than the electron thermal speed at the sheath edge ($V_{z,e} \ll v_{Te}$). In this common situation, equation~(\ref{eq:newkbohm}) reduces to 
\begin{equation}
\sum_i \frac{q_i^2}{e^2} \frac{n_{i}}{n_{e}} \frac{c_{s,i}^2}{V_{z,i}^2 - v_{T,i}^2/2} \biggl|_{z=0} \leq 1  , \label{eq:multibohm}
\end{equation}
in which $i$ label the different ion species. Equation~(\ref{eq:multibohm}) was first derived by Riemann using a fluid approach \cite{riem:95}.  The limit of equation~(\ref{eq:multibohm}) is obtained from the general Bohm criterion [equation~(\ref{eq:longbohmcrit})] by effectively asserting that deviations from Maxwellian plasmas are small and the conventional fluid theory is valid. Equation~(\ref{eq:multibohm}) can be written explicitly in terms of $f_e$ and $f_i$ by substituting equations~(\ref{eq:density}), (\ref{eq:fluidflow}) and (\ref{eq:temperature}) for $n_s$, $\vc{V}_s$ and $T_s$.  

To connect with Bohm's seminal work, consider a plasma with a single species of ions of unit charge. Then equation~(\ref{eq:multibohm}) yields 
\begin{equation}
V_{z,i} \geq \sqrt{c_{s}^2 + v_{T,i}^2 / 2}   . \label{eq:anotherbohm}
\end{equation} 
If $T_e \gg T_i$, equation~(\ref{eq:anotherbohm}) simply reduces to the usual Bohm criterion $V_{z,i} \geq c_s$. However, whereas Bohm assumed monoenergetic ions, here $V_{z,i}$ and the $T_e$ in $c_s = \sqrt{T_e/ M_i}$ are defined in terms of velocity-space moments of the ion and electron distribution functions [equations (\ref{eq:fluidflow}) and (\ref{eq:temperature})]. Explicitly in terms of $f_e$ and $f_i$, the $V_{z,i} \geq c_s$ criterion can be written
\begin{equation}
\int d^3v\, v_z f_i \geq \biggl[ \frac{1}{3} \frac{m_e}{M_i} \biggl( \int d^3v f \biggr) \biggl(\int d^3v\, v_z^2 f_e \biggr) \biggr]^{1/2}, \label{eq:kusualbohm}
\end{equation}  
in which $f$ in the $\int d^3 v f$ term can be either $f_i$ or $f_e$ because of quasineutrality at the sheath edge ($n_e \approx n_i$). 

\section{Example distributions for comparing the different Bohm criteria\label{sec:examples}}

In this section, we use ion and electron distribution functions from example presheath models discussed in previous literature to demonstrate the similarities and differences between the conventional kinetic Bohm criterion and the criterion derived in section~\ref{sec:newbohm}. The Debye length is much shorter than the ion or electron momentum transfer collision length in all of these examples. They are also concerned with single ion and electron species plasmas. Thus, we will be comparing the conventional Bohm criterion of equation (\ref{eq:kbohm}) with the single ion and electron species version of equation~(\ref{eq:longbohmcrit}) in which the densities, flow speeds and temperatures are calculated from $f_i$ and $f_e$ using equations~(\ref{eq:density}), (\ref{eq:fluidflow}) and (\ref{eq:temperature}). The example ion distribution functions considered include a monoenergetic distribution, collisionless models including the Tonks-Langmuir problem and variations of it that account for finite temperature ionization sources, as well as a Coulomb collisional presheath model where ions have a flowing Maxwellian distribution. The example electron distributions considered are Maxwellian and a collisionless model where the Maxwellian is truncated for velocities corresponding to electrons that have escaped the plasma through the sheath potential drop. 

\subsection{Monoenergetic ions and Maxwellian electrons}

The idealized plasma that Bohm considered in his original paper~\cite{bohm:49} assumed monoenergetic ions, $f_i = n_i \delta (\vc{v} - \vc{V}_i)$, and Maxwellian electrons
\begin{equation}
f_{e} = \frac{n_e}{\pi^{3/2} v_{Te}^3} \exp \biggl(- \frac{\vc{v}^2}{v_{Te}^2} \biggr) . 
\end{equation}
For these distribution functions, the components of the conventional kinetic Bohm criterion of equation~(\ref{eq:kbohm}) are $\int d^3v\, f_i/(M_i v_z^2) = n_i / (M_i V_i^2)$ and $- \int d^3v\, (\partial f_e/\partial v_z)/(M_i v_z) = n_e/T_e$. Putting these into equation~(\ref{eq:kbohm}) and applying the quasineutrality assumption gives Bohm's original criterion $V_i \geq c_s$.  Equations~(\ref{eq:stresstens}) and (\ref{eq:temperature}) give $T_i = 0$ and $\Pi_{zz,i}=0$ for monoenergetic ions. For Maxwellian electrons, $\vc{V}_i=0$, $\Pi_{zz,e} = 0$ and $d T_{e}/dz = 0$ (see section~\ref{sec:trun}). Thus, equation~(\ref{eq:multibohm}) reduces to the same condition $V_i \geq c_s$. The conventional kinetic Bohm criterion and the kinetic criterion developed in section~\ref{sec:newbohm} both reduce to the criterion of Bohm's original work~\cite{bohm:49} for monoenergetic ions and Maxwellian electrons.

\subsection{Collisionless presheath models\label{sec:coless}}

\subsubsection{Collisionless ion effects\label{sec:colesions}}

Ion kinetic effects in the presheath are often studied using hybrid models in which ions are treated with a collisionless kinetic equation, and electrons are assumed to obey a Boltzmann density profile, $n_e = n_o \exp(e \phi/T_e)$ \cite{harr:59,emme:80,biss:87a,biss:87,biss:89,riem:06,jeli:09,sher:01,robe:09,tonk:29}. The main distinguishing feature in the various models is how the ion source term is treated. The first of these theories was proposed in the seminal work of Tonks and Langmuir \cite{tonk:29}, later extended by Harrison and Thompson \cite{harr:59}. In this model ions are assumed to be born with zero energy; all of their energy being provided by subsequent acceleration through the presheath electric field.  Riemann has shown that the ion velocity distribution function at the sheath edge in the Tonks-Langmuir problem [where the ion source is singular in velocity $S_i \propto \delta(v_z)$] is independent of the ionization rate \cite{riem:06}. It satisfies the conventional Bohm criterion of equation~(\ref{eq:kbohm}) \cite{harr:59,riem:06}, which is possible because $f_i(v_z=0)=0$ at the sheath edge.

The Tonks-Langmuir problem has also been generalized to account for ion source distributions that have an energy spread. Emmert {\it et al} \cite{emme:80} assumed a source function of the flux form
\begin{equation}
S_i = h(z) \frac{v_z}{v_{Ti}^2} \exp (-v_z^2/v_{Ti}^2 ) . \label{eq:emsource}
\end{equation}
in which $h(z)$ is the spatial strength of the source. Equation~(\ref{eq:emsource}) vanishes for $v_z=0$, so the $v_z^{-1}$ moment does not diverge. A consequence of this source distribution is that the resultant ion velocity distribution function satisfies $f_i (v_z\leq0)= 0$ at the sheath edge \cite{emme:80}. Elsewhere in the presheath $f_i(v_z\leq 0) \neq 0$. Bissell has shown that Emmert's solution satisfies the conventional Bohm criterion of equation~(\ref{eq:kbohm}) \cite{biss:87}, which is only possible because $S_i(v_z=0)=0$. 

Since ions are born from ionization of neutral atoms, which themselves are expected to have a Maxwellian distribution, a natural assumption might be that the ion source function also have a Maxwellian distribution with a temperature near the neutral gas temperature. Bissell and Johnson \cite{biss:87a,biss:89} applied an analysis similar to Emmert's, but assumed a Maxwellian 
\begin{equation}
S_i = \frac{h(z)}{v_{Ti}} \exp (-v_z^2/v_{Ti}^2) .
\end{equation}
Using this Maxwellian source function, the resultant ion velocity distribution function at the sheath edge was found to have a nonzero contribution at $v_z=0$. Sheridan~\cite{sher:01} has also provided a detailed numerical study of Bissell and Johnson's model showing that $f_i(v_z=0) \neq 0$ at the sheath edge for any finite ion temperature in the source distribution. Robertson has found similar results for warmer source temperatures than Sheridan considered~\cite{robe:09}.  Similar ion velocity distribution functions with $f_i(v_z=0) \neq 0$ at the sheath edge have also been found in particle-in-cell simulations \cite{jeli:07}, and other kinetic simulations~\cite{koch:89}.   Since $f_i(v_z=0) \neq 0$ at the sheath edge for this model, the left-hand side of equation~(\ref{eq:kbohm}) diverges and the conventional kinetic Bohm criterion does not provide a useful condition limiting the ion speed at the sheath edge. However, if the Bohm criterion from section~\ref{sec:newbohm} is applied to these ion distributions, we find that the essentially fluid result from equation~(\ref{eq:anotherbohm}) holds, where the fluid variables are defined in terms of $f_i$ from equations~(\ref{eq:fluidflow}) and (\ref{eq:temperature}). 

\subsubsection{Collisionless electron effects\label{sec:trun}}

If the electron momentum transfer collision length is longer than the sheath length, the electron distribution function at the sheath edge will be depleted for velocities exceeding some threshold.  This threshold corresponds to the velocity required to overcome the sheath potential drop. Electrons that escape through the sheath are lost to the boundary surface, so the electron distribution function is depleted beyond this critical velocity by a factor of order $d/\lambda_e$, where $d$ is the distance to the boundary and $\lambda_{e}$ is the electron collision length. If the material boundary is in the $+\hat{z}$ facing direction, the truncation velocity is given by $\vc{v}_{\parallel c} = - \sqrt{2 e(|\phi_b| + \phi)/m_e}\, \hat{z}$.  Here $\phi_b$ is the potential of the boundary surface with respect to the plasma. We have chosen the plasma potential as the reference potential ($\phi_p = 0$), so $\phi$ typically takes negative values through the presheath and sheath. 

\begin{figure}
\begin{center}
\includegraphics{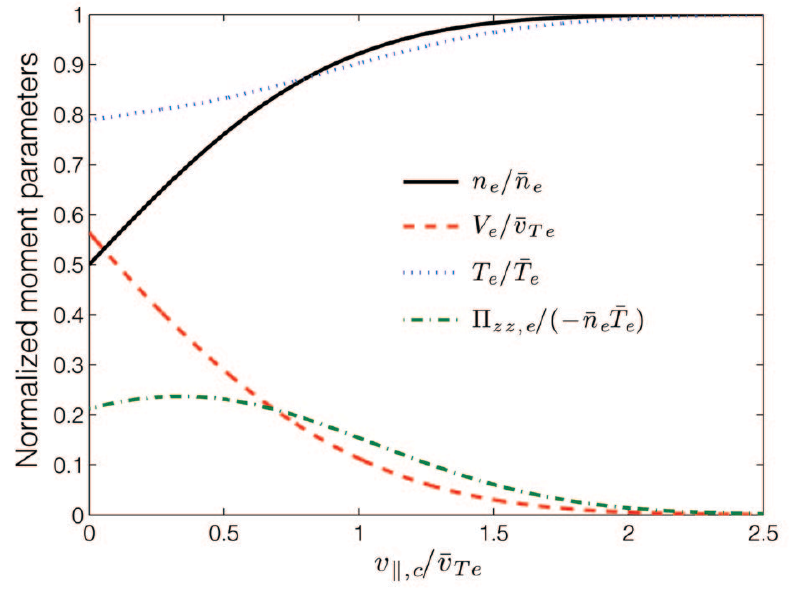}
\caption{Normalized electron density (solid black line), flow speed (dashed red line), stress (dash-dotted green line) and temperature (dotted blue line) computed from the moment equations~(\ref{eq:density}), (\ref{eq:fluidflow}), (\ref{eq:stresstens}) and (\ref{eq:temperature}) for the truncated Maxwellian distribution function of equation~(\ref{eq:truncatedm}).}
\label{fg:moments}
\end{center}
\end{figure}

Assuming that electrons are Maxwellian, aside from the truncation, the electron velocity distribution function can be written
\begin{equation}
f_e = \frac{\bar{n}_e}{\pi^{3/2} \bar{v}_{Te}^3} \exp \biggl( - \frac{v^2}{\bar{v}_{Te}^2} \biggr) H(v_z + v_{\parallel,c}) \label{eq:truncatedm}
\end{equation}
in which $H$ is the Heaviside step function, $\bar{v}_{Te} \equiv \sqrt{2 \bar{T}_e/m_e}$, and $v_{\parallel,c} \equiv |\vc{v}_{\parallel,c}|$. In terms of the fluid variable definitions from equations~(\ref{eq:density})--(\ref{eq:temperature}), the density is 
\begin{equation}
n_e = \frac{\bar{n}_e}{2} \biggl[1 + \textrm{erf} \biggl( \frac{v_{\parallel,c}}{\bar{v}_{Te}} \biggr) \biggr]  , \label{eq:edens}
\end{equation}
the flow velocity is
\begin{equation}
\vc{V}_e = \frac{1}{\sqrt{\pi}} \frac{\exp \bigl( - v_{\parallel,c}^2 / \bar{v}_{Te}^2 \bigr)}{1 + \textrm{erf} \bigl( v_{\parallel,c} / \bar{v}_{Te} \bigr)} \bar{v}_{Te}\, \hat{z}  , \label{eq:eflow}
\end{equation}
the temperature is 
\begin{equation}
T_e =\bar{T}_e \biggl\lbrace 1 - \frac{2}{3 \sqrt{\pi}} \frac{v_{\parallel,c}}{\bar{v}_{Te}} \frac{\exp \bigl(- v_{\parallel,c}^2 / \bar{v}_{Te}^2 \bigr)}{\bigl[1 + \textrm{erf} \bigl(v_{\parallel,c} / \bar{v}_{Te} \bigr) \bigr]}  - \frac{2}{3\pi} \frac{\exp \bigl( - 2 v_{\parallel,c}^2 / \bar{v}_{Te}^2 \bigr)}{\bigl[1 + \textrm{erf} \bigl(v_{\parallel,c} / \bar{v}_{Te} \bigr) \bigr]^2} \biggr\rbrace  , \label{eq:etemp}
\end{equation}
and the $\hat{z} \hat{z}$ component of the stress tensor is
\begin{equation}
\stres_{zz} = -\bar{n}_e \bar{T}_e \frac{2}{3\sqrt{\pi}} \exp \biggl(-\frac{v_{\parallel,c}^2}{\bar{v}_{Te}^2} \biggr) \biggl[ \frac{v_{\parallel,c}}{\bar{v}_{Te}} + \frac{1}{\sqrt{\pi}} \frac{\exp ( - v_{\parallel, c}^2/ \bar{v}_{Te}^2 )}{1 + \erf( v_{\parallel,c} / \bar{v}_{Te} )} \biggr]. \label{eq:estress}
\end{equation}
Equations~(\ref{eq:edens})-(\ref{eq:estress}) are shown in figure~\ref{fg:moments} as functions of $v_{\parallel,c}/\bar{v}_{Te}$.  In this problem, we will use the temperature evolution equation to close the moment equations. Thus, the heat flux, $\vc{q}_s \equiv m_s \int d^3v\, \vc{v}_r\, v_r^2 f_s/m_s$, will be required. For the distribution of equation~(\ref{eq:truncatedm}), the electron heat flux is 
\begin{equation}
q_{z,e} = n_e V_e \bar{T}_e \biggl\lbrace - \frac{1}{2} + \frac{v_{\parallel,c}^2}{\bar{v}_{Te}^2} + 3 \frac{v_{\parallel,c}}{\bar{v}_{Te}} \frac{V_{e}}{\bar{v}_{Te}} + 2 \frac{V_{e}^2}{\bar{v}_{Te}^2} \biggr\rbrace . \label{eq:eheat}
\end{equation}

\begin{figure}
\begin{center}
\includegraphics{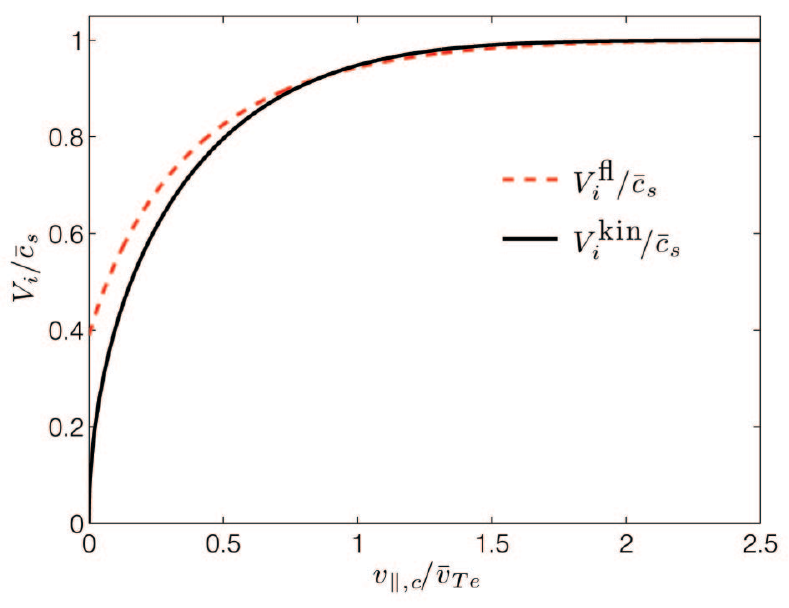}
\caption{Prediction for the minimum ion speed at the sheath edge normalized to $\bar{c}_s$ as a function of the truncation speed for the electron distribution function of equation~(\ref{eq:truncatedm}). The red dashed line shows the prediction of equation~(\ref{eq:trunmybohm}), which is an approximation that assumes electron temperature and stress gradients are negligible. The black solid line shows the prediction of equation~(\ref{eq:myebohm}), which is the full kinetic result.} 
\label{fg:eBohm}
\end{center}
\end{figure}

Figure~(\ref{fg:moments}) shows that if $v_{\parallel,c}/\bar{v}_{Te} \lesssim 1$, the $d T_e/dz$ and $d \Pi_{zz,e}/dz$ terms of equation~(\ref{eq:longbohmcrit}) may become important. 
This is a consequence of the distribution being far from Maxwellian. In this circumstance, the simplification of equation~(\ref{eq:newkbohm}) is not expected to be valid. To demonstrate that the temperature and stress gradients should not be neglected when the distribution is far from Maxwellian, we compare the predictions of equations~(\ref{eq:longbohmcrit}) and (\ref{eq:newkbohm}). Assuming that ions are monoenergetic, equation~(\ref{eq:newkbohm}) reduces to
\begin{equation}
V_i^{\flu} \geq c_s \sqrt{1 - m_e V_e^2/T_e}   .  \label{eq:ebohm}
\end{equation} 
Putting equations~(\ref{eq:edens}), (\ref{eq:eflow}) and (\ref{eq:etemp}) into (\ref{eq:ebohm}) yields the following form of a Bohm criterion 
\begin{equation}
V_i^{\flu} \geq \bar{c}_s \biggl\lbrace 1  - \frac{2}{3\sqrt{\pi}} \frac{v_{\parallel,c}}{\bar{v}_{Te}} \frac{\exp \bigl(- v_{\parallel,c}^2 / \bar{v}_{Te}^2 \bigr)}{\bigl[1 + \textrm{erf} \bigl(v_{\parallel,c} / \bar{v}_{Te} \bigr) \bigr]} - \frac{8}{3\pi} \frac{\exp \bigl( - 2 v_{\parallel,c}^2 / \bar{v}_{Te}^2 \bigr)}{\bigl[1 + \textrm{erf} \bigl(v_{\parallel,c} / \bar{v}_{Te} \bigr) \bigr]^2} \biggr\rbrace^{1/2} .\label{eq:trunmybohm}
\end{equation}
Equation~(\ref{eq:trunmybohm}) is shown as the red dashed line in figure \ref{fg:eBohm}. It is an essentially fluid result which neglects the $dT_e/dz$ and $d \Pi_{zz,e}/dz$ terms that must be determined from higher-order moments of the kinetic equation. 

A full kinetic calculation of the Bohm criterion from equation~(\ref{eq:longbohmcrit}) requires taking the spatial derivative of equations~(\ref{eq:etemp}) and (\ref{eq:estress}). At this point, both $\bar{n}_e$ and $\bar{T}_e$ in equation~(\ref{eq:truncatedm}) are expected to have a spatial dependence. A closure scheme is required to solve for the three variables $\bar{n}_e$, $V_e$ and $\bar{T}_e$. This can be provided by deriving a temperature evolution equation from the $\int d^3v\, \vc{v}_r v_r^2$ moment of the kinetic equation~(\ref{eq:pke}).  The 1-D steady-state version for electrons, neglecting collision terms, is
\begin{equation}
\frac{d}{dz} \biggl[ q_{z,e} + \biggl( \frac{5}{2} n_e T_e + \frac{1}{2} m_e n_e V_{z,e}^2 \biggr) V_{z,e} + V_{z,e} \Pi_{zz,e} \biggr] + e n_e V_{z,e} E  = 0 . \label{eq:tev}
\end{equation}
From equations~(\ref{eq:edens})--(\ref{eq:eheat}), we find that 
\begin{equation}
q_{z,e} + \biggl(\frac{5}{2} n_e T_e + \frac{1}{2} m_e n_e V_{z,e}^2 \biggr)V_{z,e} + V_{z,e} \Pi_{zz,e} = n_e V_e \bar{T}_e \biggl(2 + \frac{v_{\parallel,c}^2}{\bar{v}_{Te}^2} \biggr) .
\end{equation}
Putting this into the collisionless temperature evolution equation~(\ref{eq:tev}), and identifying the continuity equation $d(n_e V_{z,e})/dz=0$, yields
\begin{equation}
\frac{d}{dz} \biggl[ \bar{T}_e \biggl(2 + \frac{v_{\parallel,c}^2}{\bar{v}_{Te}^2} \biggr) \biggr] = -e E . \label{eq:tempint}
\end{equation}
Noting that 
\begin{equation}
\frac{d}{dz} \biggl( \frac{v_{\parallel,c}^2}{\bar{v}_{Te}^2} \biggr) = \frac{d}{dz} \biggl( \frac{e (|\phi_b| + \phi)}{\bar{T}_e} \biggr) = - \frac{e E}{\bar{T}_e} - \frac{v_{\parallel,c}^2}{\bar{v}_{Te}^2} \frac{1}{\bar{T}_e} \frac{d \bar{T}_e}{dz} ,
\end{equation}
equation~(\ref{eq:tempint}) reduces to
\begin{equation}
\frac{d \bar{T}_e}{dz} = 0  . \label{eq:tbar}
\end{equation}
Thus, in the absence of collisions, the parameter $\bar{T}_e$ is constant. The temperature $T_e$, of course, changes in space as $v_{\parallel,c}$ varies. However, it is also useful for the other examples to note that for a full Maxwellian $\bar{T}_e \rightarrow T_e$, so equation~(\ref{eq:tbar}) shows that the temperature is constant for these examples when electron collisions can be neglected.

Equation~(\ref{eq:tbar}) implies 
\begin{equation}
\frac{d}{dz} \biggl( \frac{v_{\parallel,c}}{\bar{v}_{Te}} \biggr) = - \frac{1}{2} \frac{\bar{v}_{Te}}{v_{\parallel,c}} \frac{e E}{\bar{T}_e}  .\label{eq:ddz}
\end{equation}
Using equation~(\ref{eq:ddz}) to evaluate $n_e dT_e/dz + d \Pi_{zz,e}/dz$ from equations~(\ref{eq:etemp}) and (\ref{eq:estress}), one can show
\begin{eqnarray}
- e n_e &-& \frac{1}{E} \biggl( n_e \frac{d T_e}{dz} + \frac{d \Pi_{zz,e}}{dz} \biggr) = \\ \nonumber
& & \frac{e n_e}{\bar{T}_e}  \bigl( m_e V_e^2 - T_e \bigr) \biggl(1 + \frac{\bar{v}_{Te}}{v_{\parallel,c}} \frac{\exp(-v_{\parallel,c}^2/ \bar{v}_{Te}^2)}{\sqrt{\pi} [ 1 + \erf( v_{\parallel,c} / \bar{v}_{Te})]} \biggr) .
\end{eqnarray} 
Putting this into equation~(\ref{eq:longbohmcrit}), along with the monoenergetic ion term, yields
\begin{equation}
\frac{e^2 n_i}{M_i V_i^2} - \frac{e^2 n_e}{\bar{T}_e} \biggl[ 1 + \frac{\bar{v}_{Te}}{v_{\parallel,c}} \frac{ \exp(-v_{\parallel,c}^2/ \bar{v}_{Te}^2)}{\sqrt{\pi} [ 1 + \erf (v_{\parallel,c}/ \bar{v}_{Te})]} \biggr] \leq 0  .
\end{equation}
Rearranging, the full kinetic Bohm criterion of equation~(\ref{eq:longbohmcrit}) yields 
\begin{equation}
V_i^{\kin} \geq \bar{c}_s \biggl\lbrace 1 + \frac{\bar{v}_{Te}}{v_{\parallel,c}} \frac{\exp ( - v_{\parallel, c}^2 / \bar{v}_{Te}^2)}{\sqrt{\pi} [1 + \erf(v_{\parallel,c} / \bar{v}_{Te})]} \biggr\rbrace^{-1/2}   . \label{eq:myebohm}
\end{equation} 
Equation~(\ref{eq:myebohm}) is shown as the black line in figure~\ref{fg:eBohm}. 

As expected, figure~\ref{fg:eBohm} shows that for $v_{\parallel,c}/\bar{v}_{Te} \lesssim 1$ equations~(\ref{eq:trunmybohm}) and (\ref{eq:myebohm}) give significantly different predictions. This is because the electron distribution function deviates significantly from Maxwellian in this region, and the $dT_e/dz$ and $d\Pi_{zz,e}/dz$ terms become important. For $v_{\parallel,c}/\bar{v}_{Te} \gtrsim 1$, the conventional result that ions obtain the sound speed at the sheath edge is returned. For a floating boundary, the ion and electron fluxes balance at the sheath edge, and the sheath potential drop is $|\Delta \phi_b| = T_e [1 + \ln (M_i/2\pi m_e)]/2e$. This is typically a few electron temperatures, in which case $v_{\parallel,c}/\bar{v}_{Te} > 1$. Thus, the effects of a truncated electron distribution do not significantly affect the Bohm criterion near a floating boundary. 

A simple way to construct a boundary for which $v_{\parallel,c} /v_{Te} \lesssim 1$ is to bias a probe near, or more positive than, the plasma potential. Depending on the bias, and size of the probe relative to the chamber wall, the sheath near the probe can be an ion sheath, double sheath or electron sheath~\cite{baal:07}. The parameter $v_{\parallel,c}/\bar{v}_{Te}$ can be varied using the applied potential. Figure~\ref{fg:eBohm} shows that temperature and stress gradients are important in this regime and that one must use the full kinetic result from equations~(\ref{eq:longbohmcrit}), rather than the common approximation from equation~(\ref{eq:newkbohm}). 

Next, we compare the Bohm criterion of equation~(\ref{eq:myebohm}) to that predicted by the conventional kinetic Bohm criterion of equation~(\ref{eq:kbohm}). The electron term in the conventional criterion is 
\begin{equation}
- \frac{1}{m_e} \int d^3v\, \frac{1}{v_z} \frac{\partial f_e}{\partial v_z} = \frac{n_e}{\bar{T}_e} \biggl[ 1 + \frac{\exp(-v_{\parallel,c}^2/\bar{v}_{Te}^2 )}{\sqrt{\pi}[1 + \erf(v_{\parallel,c}/\bar{v}_{Te})]} \frac{\bar{v}_{Te}}{v_{\parallel,c}} \biggr] . \label{eq:okbohm}
\end{equation}
Applying equation~(\ref{eq:okbohm}) and that ions are monoenergetic to equation~(\ref{eq:kbohm}) yields
\begin{equation}
V_i \geq \bar{c}_s \biggl\lbrace 1 +  \frac{\bar{v}_{Te}}{v_{\parallel,c}} \frac{\exp(-v_{\parallel,c}^2/\bar{v}_{Te}^2 )}{\sqrt{\pi}[1 + \erf(v_{\parallel,c}/\bar{v}_{Te})]} \biggr\rbrace^{-1/2}   . \label{eq:trunkbohm}
\end{equation} 
This is the same as equation~(\ref{eq:myebohm}), which was obtained from the method of positive-velocity moments. 

In this example, the conventional kinetic Bohm criterion and the criterion from section~\ref{sec:newbohm} gave the same result because the electron distribution chosen is a possible solution of the Vlasov equation (it satisfies $\partial f_e/\partial v_z|_{v_z=0} = 0$). However, if the distribution function of equation~(\ref{eq:truncatedm}) is modified slightly, the conventional result can change dramatically. For example, if the electrons have a small drift velocity in the $\hat{z}$ direction [so that $v^2 \rightarrow (\vc{v} - \bar{\vc{V}}_{z,e})^2$ in the exponential of equation~(\ref{eq:truncatedm})], then equation~(\ref{eq:kbohm}) leads to the prediction $1/(M_i V_i^2) \leq - \infty$. With the new method from section~\ref{sec:newbohm}, adding a small drift leads to $\mathcal{O}(\bar{V}_e/\bar{v}_{Te})$ corrections to equation~(\ref{eq:myebohm}). These are negligible as long as the applied drift is much slower than the electron thermal speed.

\subsection{Collisional presheath\label{sec:colps}}

Finally, we consider plasma in which the ion-ion and electron-electron collision lengths ($\lambda^{i-i}$ and $\lambda^{e-e}$) are much shorter than the presheath length (which is typically the ion-neutral collision length). In this situation, the ion distribution function at the sheath edge is a flow-shifted Maxwellian. If the electron-electron collision length is longer than the sheath thickness, the electron distribution function at the sheath edge will be truncated at a velocity corresponding to the sheath energy. However, we found in section~\ref{sec:trun} that this truncation rarely affects the Bohm criterion. Thus, we assume that electrons have a stationary Maxwellian distribution. 

For flowing Maxwellian ions and stationary Maxwellian electrons, the Bohm criterion from equation~(\ref{eq:newkbohm}) simply reduces to $V_i \geq \sqrt{c_s^2 + v_{Ti}^2/2}$. However, the ion term in the conventional kinetic Bohm criterion from equation~(\ref{eq:kbohm}) diverges for a flowing Maxwellian distribution
\begin{equation}
\frac{1}{M_i} \int d^3v \frac{f_{Mi} (\vc{v})}{v_z^2} = \frac{n_i}{2 \sqrt{\pi} T_i} \int_{-\infty}^\infty dv_z \frac{\exp(-v_z^2 / v_{Ti}^2)}{v_z^2} \rightarrow \infty 
\end{equation}
[see equation~(\ref{eq:integrationbp})]. Thus, equation~(\ref{eq:kbohm}) leads to an incorrect statement, $\infty \leq n_e/T_e$, whereas equation~(\ref{eq:newkbohm}) gives a criterion identical to that provided by conventional fluid theory. In section~\ref{sec:collisions} we show that collisional presheaths ($\lambda^{i-i}, \lambda^{e-e} \ll l$) are common in laboratory plasmas if the electrons are much hotter than ions. 

\section{Coulomb collisions in the presheath\label{sec:collisions}}

In this section, the role of Coulomb collisions in the presheath is discussed. We are interested in determining when the ion-ion collision length is shorter than the presheath length. If this condition is met, the ion distribution function is expected to be a flowing Maxwellian at the sheath edge. This situation provides an example that is commonly found in laboratory plasmas, but where the conventional Bohm criterion and the one derived in section~\ref{sec:bohmcriterion} provide very different predictions.

Coulomb collisions in a plasma are described by the collision operator of equation~(\ref{eq:collop}). If the plasma is stable, Lenard \cite{lena:60} and Balescu \cite{bale:60} showed that the appropriate collisional kernel, $\Q$, is equation~(\ref{eq:qlb}). Recently, we generalized Lenard-Balescu theory to also account for unstable plasmas \cite{baal:08,baal:10}. For an unstable plasma, the collisional kernel is the sum of the stable plasma (Lenard-Balescu) term and an instability-enhanced collision term: $\Q = \Q_{\LB} + \Q_{\IE}$. The instability-enhanced collisional kernel is \cite{baal:08,baal:10}
\begin{eqnarray}
\Q_{\IE} &=& \frac{2 q_s^2 q_{s^\prime}^2}{m_s} \int d^3k \frac{\vc{k} \vc{k}}{k^4} \sum_j \frac{\gamma_j}{(\omega_{R,j} - \vc{k} \cdot \vc{v})^2 + \gamma_j^2}  \label{eq:qie} \\ \nonumber
& \times & \frac{\exp(2 \gamma_j t)}{[(\omega_{R,j} - \vc{k} \cdot \vc{v}^\prime)^2 + \gamma_j^2] | \partial \hat{\varepsilon} (\vc{k}, \omega)/ \partial \omega |_{\omega_j}^2} , 
\end{eqnarray}
in which $\omega_{R,j}$ is the real part of the angular frequency and $\gamma_j$ the imaginary part of the angular frequency of the $j^{th}$ unstable mode. Time in equation~(\ref{eq:qie}) is calculated in the rest frame of the unstable mode. For convective instabilities, $t$ can be translated into distance in the laboratory frame \cite{baal:08}. We will be considering ion-acoustic instabilities here, which are convective.

It has been shown that both the Lenard-Balescu~\cite{lena:60,mont:64} and instability-enhanced~\cite{baal:10} terms cause component collision operators to evolve to uniquely Maxwellian distribution functions. Thus, the distance over which a non-Maxwellian distribution function evolves to Maxwellian is determined by the shorter of the stable plasma collision length $\lambda_{\LB}^{s-s}\approx \bar{v}/\nu_{\LB}^{s-s}$ or the instability-enhanced collision length $\lambda_{\IE}^{s-s} \approx \bar{v}/\nu_{\IE}^{s-s}$. Here $\nu$ is the collision frequency, $s$ is the particular species considered, $s-s$ refers to self collisions within this species, and $\bar{v}$ is the velocity of the test particle considered in the rest frame of the $s$ distribution (i.e., it is how far in velocity space a test particle must go to be thermalized, which is approximately $|\vc{v} - \vc{V}_s|$). We are interested in like-particle collisions $i-i$ and $e-e$, but not unlike-particle collisions $e-i$ or $i-e$ because they are much less frequent. 

The collision frequency can be estimated directly from the collision operator: $\nu^{s-s} \sim C(f_s, f_s) / f_s$. Using $\partial /\partial \vc{v} \sim \bar{v}^{-1}$ in equation~(\ref{eq:collop}) shows that the collision frequency also consists of the sum of stable plasma and instability-enhanced terms
\begin{equation}
\nu^{s-s} \sim \frac{C(f_s,f_s)}{f_s} \sim \frac{n_s}{m_s \bar{v}^2} \bigl( \Q_{\LB}^{s-s} + \Q_{\IE}^{s-s} \bigr)  . \label{eq:nueqn}
\end{equation} 

We first consider the stable plasma contribution, $\nu^{s-s}_{\LB} \sim n_s Q_{\LB}^{s-s} / (m_s \bar{v}^2)$, which requires evaluating $\Q_{\LB}^{s-s}$ from equation~(\ref{eq:qlb}). Since the dielectric function at $\omega = \vc{k} \cdot \vc{v}$ is adiabatic over most of the $k$-space integral, $\hat{\varepsilon}(\vc{k}, \vc{k} \cdot \vc{v}) \approx 1 + k^{-2} \lambda_{De}^{-2}$, the Lenard-Balescu kernel reduces to the Landau collisional kernel \cite{land:36}
\begin{equation}
\Q_{L} = \frac{2 \pi q_s^2 q_{s^\prime}^2}{m_s} \frac{u^2 \ident - \vc{u} \vc{u}}{u^3} \ln \Lambda  , \label{eq:landau}
\end{equation}
in which $\vc{u} \equiv \vc{v} - \vc{v}^\prime$ and $\Lambda \approx 12 \pi n_e \lambda_{De}^3$.   Taking $\vc{u} \sim \bar{v}$, the dominant term of the stable plasma collisional kernel for like-particle collisions is $Q_{\LB}^{s-s} \approx 2\pi q_s^4 \ln \Lambda /(m_s \bar{v})$, and the stable plasma collision frequency is  
\begin{equation}
\nu_{\LB}^{s-s} \approx \frac{2\pi n_s q_s^4}{m_s^2 \bar{v}^3} \ln \Lambda \label{eq:nuLB}.
\end{equation} 

Next, we consider the possibility of ion-acoustic instabilities enhancing collisions in the presheath. Ion-acoustic waves have a phase speed that is slow compared to the electron thermal speed, $\omega/kv_{Te} \ll 1$ and fast compared to the ion thermal speed $(\omega - \vc{k} \cdot \vc{V}_i)/kv_{Ti} \gg 1$. We assume that electrons are much hotter than ions, $T_i/T_e \ll 1$, in which case ion Landau damping can be neglected. Under these assumptions, the electrostatic dielectric function of equation~(\ref{eq:dielec}) reduces to 
\begin{equation}
\hat{\varepsilon} = 1 + \frac{1}{k^2 \lambda_{De}^2} - \frac{\omega_{pi}^2}{(\omega - \vc{k} \cdot \vc{V}_i)^2} + i \frac{\sqrt{\pi}}{k^2 \lambda_{De}^2} \frac{\omega}{k v_{Te}} .  \label{eq:reducedd}
\end{equation} 
In obtaining equation~(\ref{eq:reducedd}), we have assumed that the electron distribution is Maxwellian. Accounting for a truncated Maxwellian of the form of equation~(\ref{eq:truncatedm}) leads to negligibly small corrections as long as $v_{\parallel,c}/v_{Te} \gtrsim 1$. A dielectric function for the truncated Maxwellian can be developed using the incomplete plasma dispersion function \cite{fran:71}, which shows that corrections to equation~(\ref{eq:reducedd}) are of order $\mathcal{O}\lbrace \exp(-v_{\parallel,c}^2/v_{Te}^2) v_{Te}/v_{\parallel,c} \rbrace$. We assume these are negligible here.

\begin{figure}
\begin{center}
\includegraphics{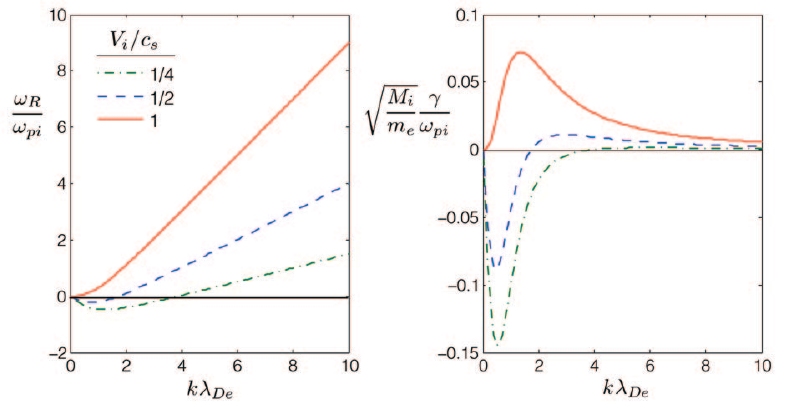}
\caption{Real and imaginary parts of the ion-acoustic dispersion relation, equation~(\ref{eq:disp}), for three values of the ion fluid speed (normalized to the sound speed) that are found in presheaths: 1/4 (green, dash-dotted line), 1/2 (blue, dashed line) and 1 (red, solid line). We have used $\vc{k} \cdot \vc{V}_i \approx k V_i$ in these plots.}
\label{fg:disp_fig}
\end{center}
\end{figure}

Solving for the roots of equation~(\ref{eq:reducedd}) gives the ion-acoustic dispersion relation 
\begin{equation}
\omega_{\pm} = \biggl( \vc{k} \cdot \vc{V}_i \pm \sqrt{\frac{n_i}{n_e}} \frac{k c_s}{\sqrt{1 + k^2 \lambda_{De}^2}} \biggr) \biggl(1 \mp i \sqrt{\frac{n_i}{n_e}} \frac{\sqrt{\pi m_e / 8 M_i}}{(1 + k^2 \lambda_{De}^2)^{3/2}} \biggr) . \label{eq:disp}
\end{equation}
A growing wave is present as long as the ion fluid speed is large enough: $| \vc{k} \cdot \vc{V}_i| > k c_s \sqrt{n_i} /\sqrt{n_e(1 + k^2 \lambda_{De}^2)}$. Equation~(\ref{eq:disp}) is plotted in figure~\ref{fg:disp_fig} for three representative values of the ion fluid speed in the presheath assuming the plasma is neutral ($n_i = n_e$). Figure~\ref{fg:disp_fig} shows that the relevant wavelengths of unstable modes are near the electron Debye length (or shorter). 

\ref{app:ie} provides a calculation of the instability-enhanced collisional kernel when ion-acoustic instabilities are present. Putting this result from equation~(\ref{eq:qiefinal}) into equation~(\ref{eq:nueqn}), we find that the instability-enhanced collision frequency for like-particle collisions is 
\begin{equation}
\nu_{\IE}^{s-s} \approx \frac{\nu_{\LB}^{s-s}}{8 \ln \Lambda} \frac{1 + \kappa_c^2}{(1 + \kappa_c^2)^2} \exp \biggl(\sqrt{\frac{\pi m_e n_i}{16 M_i n_e}} \frac{Z}{\lambda_{De}} \biggr)  \label{eq:nuie}
\end{equation}
when ion-acoustic instabilities are present. In equation~(\ref{eq:nuie}), $\nu_{\LB}^{s-s}$ is the stable plasma collision frequency from equation~(\ref{eq:nuLB}) and $\kappa_c$, defined in equation~(\ref{eq:kappac}), is a parameter that accounts for the fraction of $k$-space that is unstable. Here $Z$ is a spatial coordinate alined with $z$, but is shifted so that its origin corresponds to the location where the first unstable wave is excited. For the presheath, $Z=0$ is at the presheath-plasma interface, and $Z$ takes positive values through the presheath. 

Finally, we apply equations~(\ref{eq:nuLB}) and (\ref{eq:nuie}) to the parameters of an experiment in which Claire {\it et al} \cite{clai:06} used laser induced fluorescence to diagnose the ion velocity distribution function throughout the presheath. Similar experiments have also been conducted by others \cite{bach:95,oksu:01}. Claire {\it et al} studied an argon discharge with plasma parameters $T_i = 0.027$ eV, $T_e = 2.5$ eV and $n = 5.5 \times 10^9$ cm$^{-3}$ at a neutral pressure of $1.8 \times 10^{-4}$ mbar.  They found that the ion velocity distribution function was Maxwellian in the entrance (bulk plasma facing) region of the presheath, had a three-part distribution in the middle of the presheath that was qualitatively similar to collisionless presheath models such as those discussed in section~\ref{sec:coless} (they compared to Emmert's \cite{emme:80} model in particular), and became Maxwellian again near the sheath edge. The finding that the ion velocity distribution function became Maxwellian between the mid presheath and the sheath edge was purported to be a new and surprising result that is not predicted by previous kinetic theories of the presheath. Here we show that this measurement can be explained by ion-ion collisions enhanced by convective ion-acoustic instabilities. 

Consider the stable plasma contribution to the ion-ion collision length, $\lambda^{i-i}_{\LB} \approx \bar{v}/\nu_{\LB}^{i-i}$, using equation~(\ref{eq:nuLB}).  For the parameters of this experiment, $\lambda^{i-i}_{\LB} = 1.8 \times 10^{-13}\, \bar{v}^4$~m, in which $\bar{v}$ is in m/s. Near the entrance to the presheath, the ion flow speed is very small and the relevant speed is the ion thermal speed $\bar{v} \approx v_{Ti}$. Putting in $\bar{v} = v_{Ti} = 3.4 \times 10^2$ m/s gives $\lambda_{\LB}^{i-i} = 2.4$ mm. Near the presheath-sheath boundary, the ion flow speed is nearly the sound speed, which is much larger than the thermal speed, so $\bar{v} \approx c_s$. Using $\bar{v} = c_s = 2.4 \times 10^3$~m/s gives $\lambda_{\LB}^{i-i} = 6.0$~m. The presheath in these argon discharges is a couple of times the ion-neutral collision length \cite{oksu:02,oksu:05}. We consider the presheath to be collisional for ion-ion collisions if the ion-ion collision length is smaller than the ion-neutral collision length: $\lambda^{i-i}/\lambda^{i-n} < 1$. If this is the case, ion-ion collisions are frequent in the presheath and the ion distribution function should be a flow-shifted Maxwellian. At $1.8 \times 10^{-4}$ mbar, the neutral density is $n_n = 4.3 \times 10^{12}$ cm$^{-3}$. The total ion-neutral collision cross section for the energies of interest is approximately $\sigma \approx 1 \times 10^{-14}$~cm$^{2}$ \cite{lieb:05}. Thus, the ion-neutral collision length is $\lambda^{i-n} \approx 1/(n_n \sigma) = 23$~cm. According to the stable plasma contribution to the collision frequency, ions near the entrance to the presheath are collisional, but ions near the presheath-sheath boundary are not.  

\begin{figure}
\begin{center}
\includegraphics{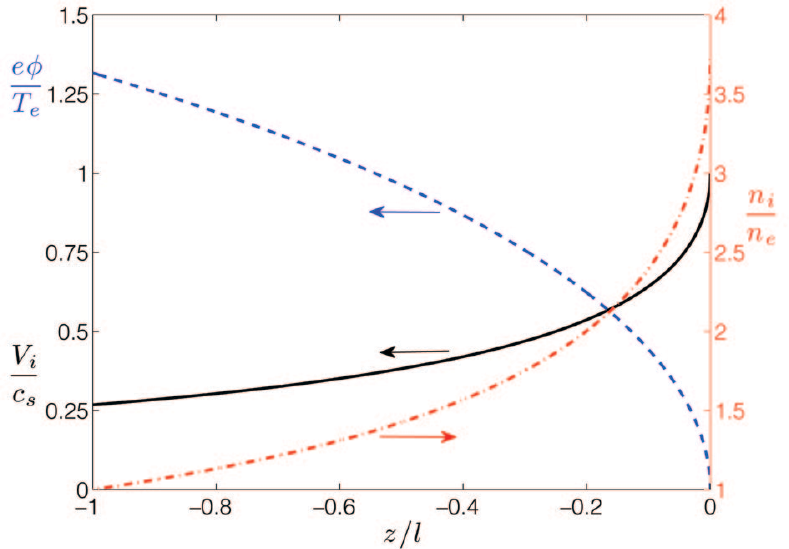}
\caption{The electrostatic potential (dashed blue line) and ion fluid flow (solid black line) throughout the presheath using the modified mobility limited flow model (left axis). Also shown on the right axis is the estimated ion to electron density ratio $n_i/n_e$ (dotted red line).}
\label{fg:presheath}
\end{center}
\end{figure}

An estimate of ion collisionality throughout the presheath can be made using a model for the ion flow speed. For this estimate, we apply a well known modified-mobility-limited-flow model from Riemann \cite{riem:97}
\begin{equation}
\frac{e \phi}{T_e} = \ln \biggl(\frac{c_s}{V_i} \biggr) \ \ \ \textrm{and} \ \ \ dz = dV_i \biggl( \frac{c_s^2 - V_i^2}{V_i^2\, \nu^{i-n}} \biggr) .\label{eq:presheath}
\end{equation}
Two models for the ion-neutral collision length are commonly used: constant collision length $\lambda^{i-n} = l$, $\nu^{i-n} = V_i/l$, or constant collision frequency $\nu^{i-n} \approx c_s/l$. For simplicity, we apply the constant collision frequency model.  The model of equation~(\ref{eq:presheath}) has been verified experimentally to a distance $z \approx 2 \lambda^{i-n}$ \cite{oksu:02,oksu:05}. We choose a coordinate system where $z=0$ is the sheath edge, and $z$ takes negative values throughout the presheath to a distance $z=-l = -2 \lambda^{i-n}$. For the constant ion-neutral collision frequency, equation~(\ref{eq:presheath}) yields $e\phi/T_e = \textrm{arccosh}(1-z/l)$ and 
\begin{equation}
V_i = c_s \biggl[1 - \frac{z}{l} \bigl(1 - \sqrt{1 - 2l/z} \bigr) \biggr]. \label{eq:vps}
\end{equation}
The profiles for electrostatic potential and ion flow speed through the presheath are shown in figure~\ref{fg:presheath}.

\begin{figure}
\begin{center}
\includegraphics{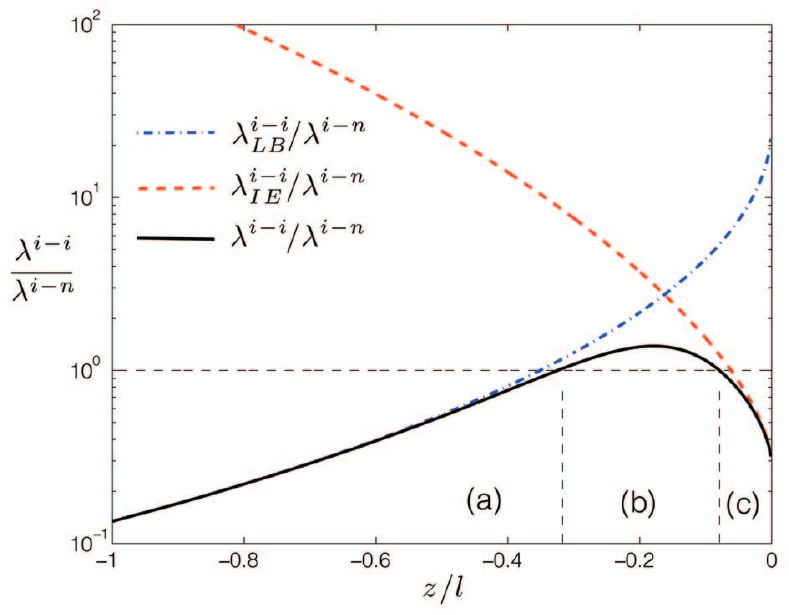}
\caption{Ion-ion collision frequencies normalized to the ion-neutral collision frequency through the presheath using the experimental parameters of Claire \etal~\cite{clai:06}. Shown are the stable plasma contribution $\lambda_{\LB}^{i-i}$ (dash-dotted blue line), the instability-enhanced contribution $\lambda_{\IE}^{i-i}$ (dashed red line) and the total collision length (solid black line). Ions are considered collisional when the total collision length is less than 1 on this scale.}
\label{fg:icoll}
\end{center}
\end{figure}

Using $\bar{v} = V_i$, the stable plasma contribution to the ion-ion collision length is plotted in figure~\ref{fg:icoll}. We use the criterion that the plasma is collisional if $\lambda^{i-i}/\lambda^{i-n}<1$. Figure~\ref{fg:icoll} shows that if one considers only the stable plasma contribution to the ion-ion collision length, ions from the entrance to the presheath until about half-way through it are collisional. Thus, the ion distribution function in this entrance region should be Maxwellian, which agrees with the measurements in Claire \etal \cite{clai:06}. However, as the sheath is approached, the ion flow speed increases significantly and the ion-ion collision length, which is $\propto \bar{v}^4$, becomes much longer than the ion-neutral collision length. This suggests that ions in the mid presheath to the sheath edge are nearly collisionless. Thus, the stable plasma contribution to ion-ion collisions alone cannot explain the finding by Claire \etal that the ion distribution function becomes Maxwellian near the sheath edge. This suggests that some other mechanism for ion-ion collisions must be present.

The apparently anomalous ion-ion scattering near the sheath edge in Claire \etal \cite{clai:06} can be explained by the presence of ion-acoustic instabilities. The instability-enhanced collision frequency can be estimated from equation~(\ref{eq:nuie}) by inputting an ion flow speed and ion-to-electron density ratio through the presheath. We use equation~(\ref{eq:vps}) for the ion flow model, which determines $\kappa_c$ through equation~(\ref{eq:kappac}). Estimating the density ratio is a bit more difficult because the model in equation~(\ref{eq:presheath}) takes $n_i=n_e$ at the sheath edge, but has $n_i \neq n_e$ into the plasma. Physically, of course, the bulk plasma is neutral and neutrality is broken as the sheath is approached. We take for an estimate $n_i/n_e \approx 1/\exp\lbrace e[\phi - \phi(z=-l)]/T_e\rbrace$, which assumes that the ion density is approximately constant and that the electron density has a Boltzmann drop. The resultant density ratio is shown in figure~\ref{fg:presheath}. Using these flow and density profiles, the instability-enhanced ion-ion collision length is shown in figure~\ref{fg:icoll}. 

\begin{figure}
\begin{center}
\includegraphics{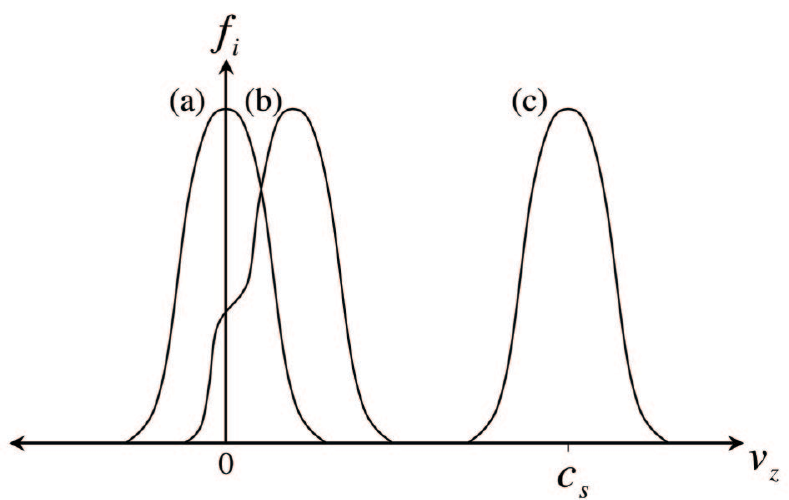}
\caption{Sketch of the predicted ion velocity distribution function at different locations in the presheath. This shows (a) a stationary Maxwellian distribution, (b) a three-region distribution predicted by collisionless models of the presheath (see section~\ref{sec:colesions}), and (c) a flowing Maxwellian distribution. Distribution (a) is expected in the plasma-facing region of the presheath where stable plasma ion-ion collisions are frequent, (b) is expected in a mid-presheath region where ions are nearly collisionless, and (c) is expected near the sheath-presheath boundary where ion-acoustic instabilities enhance ion-ion collisions causing them to be collisional again. These correspond to the expected distributions in regions (a), (b) and (c) of figure~\ref{fg:icoll}.}
\label{fg:fifig}
\end{center}
\end{figure}

Figure~\ref{fg:icoll} shows that ion-acoustic instabilities determine the ion-ion collision length near the sheath edge, shrinking it by nearly two orders of magnitude. For most of the presheath, the ion-acoustic instabilities do not enhance collisions, which are dominated by the stable plasma rate. The result is a presheath with three regions: (a) The entrance region is collisional because of conventional stable plasma collisions. As the ion flow speed increases, these become less significant because $\lambda_{\LB}^{i-i} \propto \bar{v}^4$. (b) Ions then become collisionless for a region in the mid presheath. (c) Closer to the sheath edge, ion-acoustic instabilities enhance ion-ion collisions and ions once again become collisional. Thus, the predictions of this model for the plasma parameters of \cite{clai:06} are that ions are Maxwellian in both the plasma-facing entrance (a) and sheath-facing exit (c) regions of the presheath. In the mid-presheath (b), ions should have a distribution characteristic of the collisionless models from section~(\ref{sec:colesions}). Each of these three regions can be identified in the measurements of Claire~\etal~\cite{clai:06}. The predicted ion distribution functions are shown schematically in figure~\ref{fg:fifig}. A proof that instability-enhanced collisions cause ions (and electrons) to evolve to a unique Maxwellian distribution is shown in reference~\cite{baal:10}. This property holds as long as $\gamma_j^2/\omega_{R,j}^2 \ll 1$. Here $\gamma_j^2/\omega_{R,j}^2 \sim m_e/M_i =1.3 \times 10^{-5}$.

Ion-acoustic instabilities not only enhance ion-ion collisions, but also electron-electron collisions. Section~\ref{sec:trun} discussed that if electrons are collisionless in the presheath, an otherwise Maxwellian distribution function will be truncated beyond a certain velocity corresponding to electrons that traverse the sheath and are lost from the plasma. This truncated Maxwellian has the form of equation~(\ref{eq:truncatedm}). However, if $\lambda^{e-e} \lesssim l$, electron-electron collisions become frequent enough that they cause this distribution to become Maxwellian again. Because momentum is conserved in these electron-electron collisions, the resultant Maxwellian will have a slight flow shift towards the sheath. The stable plasma contribution to the electron-electron collision length for thermal particles is $\lambda_{\LB}^{e-e}(\bar{v} = v_{Te}) = 6.0$ m [note that $\lambda^{i-i}(\bar{v} =c_s) = \lambda^{e-e} (\bar{v} = v_{Te})$]. This is much longer than the presheath length. The stable plasma collision frequency alone leads to the prediction that electrons are collisionless in the presheath and should have a truncated Maxwellian distribution. 

Claire~\etal~\cite{clai:06} does not present measurements of the electron distribution function in their discharge. However, others have measured the electron distribution in similar discharges and found it to be Maxwellian, even beyond the energy at which depletion should be expected. The first, and most famous, of these measurements was taken by Langmuir in 1925~\cite{lang:25}. This apparently anomalous electron-electron scattering near the plasma boundaries was later named ``Langmuir's paradox''\cite{gabo:55}. We recently showed that Langmuir's measurement can be understood through the mechanism of instability-enhanced collisions from ion-acoustic instabilities~\cite{baal:09a}. The same physics applies to the discharge parameters of Claire~\etal. Noting that $\lambda_{\IE}^{i-i}(\bar{v}=c_s) = \lambda_{\IE}^{e-e}(\bar{v} = v_{Te})$, the relevant ion-ion and electron-electron collision lengths are the same at the sheath edge. Thus, ion-acoustic instabilities enhance electron-electron collisions near the presheath as well as ion-ion collisions. These are frequent enough to expect a Maxwellian electron distribution function close to the sheath~\cite{baal:09a}. It is also noteworthy that in the sheath itself $n_i/n_e$ becomes quite large, which exponentially enhances the instability-enhanced collisions in equation~(\ref{eq:nuie}). The sheath itself, although spatially narrow, is also collisional for ion-ion and electron-electron collisions. 

\section{Conclusions\label{sec:conclusions}}

The central conclusion of this work is that the conventional kinetic Bohm criterion of equation~(\ref{eq:kbohm}) places undue importance on the low-velocity portion of the ion and electron distribution functions. Theoretical literature has generally accepted equation~(\ref{eq:kbohm}) along with the consequence that slow ions dominate the Bohm criterion~\cite{riem:06}. This equation has been used despite the fact that whenever $f_i(v_z=0) =0$, or $\partial f_e/\partial v_z|_{v_z=0}=0$ at the sheath edge, a divergent integral arrises that renders equation~(\ref{eq:kbohm}) unusable.  In section~\ref{sec:examples}, we considered several common example plasmas in which this occurs. 

We showed in section~\ref{sec:deficiencies} that the primary misstep in derivations of equation~(\ref{eq:kbohm}) is the neglect of a collision operator term of the form $\int d^3v\, C(f_s)/v_z$. Although collisions are typically negligible, $C(f_s)|_{v_z=0} \neq 0$ when $f_s(v_z=0)\neq 0$, or $\partial f_s/\partial v_z|_{v_z=0}\neq 0$. Thus, the neglected $v_z^{-1}$ moment of the collision operator actually diverges for precisely the same distribution functions that cause equation~(\ref{eq:kbohm}) to diverge. This term is necessary in order to make sense of the divergences that arise in the conventional kinetic Bohm criterion.  It was also shown that the ion term of equation~(\ref{eq:kbohm}) is based on a misapplication of integration-by-parts if a general ion distribution function is to be allowed. 

An alternative kinetic Bohm criterion based upon positive-exponent velocity moments of the plasma kinetic equation was derived in section~\ref{sec:newbohm}. The result is similar to conventional fluid theories, but where the fluid variables are defined in terms of positive-exponent velocity moments of the distribution functions. Developing the theory in this manner avoids the possibility of divergent integrals. In contrast to the conventional kinetic Bohm criterion, slow ions and electrons have no particular significance in this new kinetic formulation of the Bohm criterion. This model was compared with equation~(\ref{eq:kbohm}) for several example distribution functions that are common in various laboratory plasmas. Equation~(\ref{eq:kbohm}) contained divergent integrals for some of these examples, while the new theory provided the condition that the ion flow speed be supersonic at the sheath edge. 

In section~\ref{sec:collisions}, we considered the role of Coulomb collisions in determining the ion distribution function in the presheath. Typically ions are assumed to be either collisionless or to collide only with neutrals in the presheath because the ion-ion collision length in a stable plasma is usually much longer than the presheath length near the sheath edge. However, the presheath of plasmas in which $T_i/T_e \ll 1$ can be unstable to ion-acoustic instabilities. Section~\ref{sec:collisions} shows that these ion-acoustic instabilities shorten the ion-ion collision length to such a degree that ions can be considered collisional (to ion-ion collisions) near the presheath-sheath boundary and through the sheath. When this happens, the ions have a flowing Maxwellian distribution function at the sheath edge. A flowing Maxwellian distribution is one example where the conventional kinetic Bohm criterion of equation~(\ref{eq:kbohm}) diverges. The kinetic Bohm criterion of section~\ref{sec:newbohm} simply reduces to the fluid-like result of a supersonic ion flow speed for this case. 

\ack 
This material is based upon work supported under a National Science Foundation Graduate Research Fellowship (S.D.B.) and by the U.S. Department of Energy under grant No. DE-FG02-86ER53218. 

\appendix
\section{Ion-acoustic instability-enhanced collisions\label{app:ie}}
\setcounter{section}{1}

Since the ion-acoustic instabilities of equation~(\ref{eq:disp}) satisfy $\gamma_j / \omega_{R,j} \ll 1$, the instability-enhanced collisional kernel can be approximated by \cite{baal:10}
\begin{equation}
\Q_{\IE}^{s-s} \approx \frac{2\pi q_s^4}{m_s} \int d^3k \frac{\vc{k} \vc{k}}{k^4} \frac{\delta [\vc{k} \cdot (\vc{v} - \vc{v}^\prime)]\, \delta(\omega_{R,j} - \vc{k} \cdot \vc{v}) \exp(2 \gamma_j t)}{\gamma_j\, \bigl| \partial \hat{\varepsilon}(\vc{k}, \omega) / \partial \omega \bigl|_{\omega_j}^2}  . \label{eq:qaprox}
\end{equation}
Corrections to this approximation are of order $\mathcal{O}(\gamma_j^2/\omega_{R,j}^2) \sim m_e/M_i \ll 1$. From equation~(\ref{eq:reducedd}), we note that 
\begin{equation}
\frac{\partial \hat{\varepsilon}}{\partial \omega} \biggl|_{\omega_j} \approx \frac{2 \omega_{pi}^2}{(\omega_j - \vc{k} \cdot \vc{V}_i)^3} . \label{eq:epapprox}
\end{equation}
Applying the dispersion relation of equation~(\ref{eq:disp}), $\omega_j \approx \vc{k} \cdot \vc{V}_i - k c_s \sqrt{n_i} / \sqrt{n_e(1 + k^2 \lambda_{De}^2)}$, to equation~(\ref{eq:epapprox}) gives
\begin{equation}
\biggl| \frac{\partial \hat{\varepsilon}}{\partial \omega} \biggr|_{\omega_j}^2 = \frac{4}{k^2 c_s^2} \frac{n_e}{n_i} \frac{(1 + k^2 \lambda_{De}^2)^3}{k^4 \lambda_{De}^4}   . \label{eq:epmid}
\end{equation}

The second delta function in equation~(\ref{eq:qaprox}) can be estimated from the more elementary form written as a Lorentzian 
\begin{equation}
\delta(\omega_{R,j} - \vc{k} \cdot \vc{v}) \approx \frac{1}{\pi} \frac{\gamma_j}{(\omega_{R,j} - \vc{k} \cdot \vc{v})^2 + \gamma_j^2} \approx \frac{1}{\pi} \frac{n_e}{n_i} \frac{\gamma_j}{k^2 c_s^2}  . \label{eq:deltamid}
\end{equation}
Putting equations~(\ref{eq:epmid}) and (\ref{eq:deltamid}) into (\ref{eq:qaprox}) yields 
\begin{equation}
\Q_{\IE}^{s-s} \approx \frac{1}{2} \frac{q_s^4}{m_s} \int d^3k \frac{\vc{k} \vc{k}}{k^4} \delta [\vc{k} \cdot (\vc{v} - \vc{v}^\prime)] \frac{k^4 \lambda_{De}^4}{(1 + k^2 \lambda_{De}^2)^3} e^{2 \gamma t}  . \label{eq:midqie}
\end{equation} 

Next, we evaluate $2 \gamma t$ for the convective ion-acoustic waves. Reference \cite{baal:08} shows that the $\exp \bigl(2 \gamma t)$ term in equation~(\ref{eq:midqie}) must be calculated in the rest frame of the unstable mode.  Since the ion-acoustic instability is convective, 
\begin{equation}
2 \gamma t = 2 \int_{\vc{x}_o(\vc{k})}^{\vc{x}} d \vc{x}^\prime \cdot \frac{\vc{v}_g \gamma}{| \vc{v}_g|^2} \label{eq:2gamtnew}
\end{equation} 
in which $\vc{v}_g \equiv \partial \omega_R / \partial \vc{k}$ is the group velocity, $\vc{x}_o (\vc{k})$ is the location in space where wavevector $\vc{k}$ becomes unstable, and the integral $d \vc{x}^\prime$ is taken along the path of the mode.  In principle, the spatial integral in equation~(\ref{eq:2gamtnew}) requires integrating the profile of $\gamma$ and $\vc{v}_g$, which change through the presheath due to variations in the ion fluid speed and the electron density. It also requires knowing the spatial location $\vc{x}_o(\vc{k})$ at which each wavevector $\vc{k}$ becomes excited. In estimating equation~(\ref{eq:2gamtnew}), we assume that changes from spatial variations are weak, and we account for $\vc{x}_o (\vc{k})$ by only integrating over the unstable $\vc{k}$ for each spatial location $\vc{x}$. Following these approximations we obtain
\begin{equation}
2 \gamma t \approx \frac{2 Z \gamma}{v_g} , \label{eq:2gamtapprox} 
\end{equation}
in which $Z$ is a shifted coordinate (with respect to $z$) that takes as its origin the location where the first instability onset occurs. In this case, $Z=0$ will be the presheath-plasma boundary. The group speed is approximately the phase speed ($v_g \approx \omega_{R,j}/k$) for the ion-acoustic waves, so 
\begin{equation}
2 \gamma t \approx \sqrt{\frac{\pi m_e n_i}{2 M_i n_e}} \frac{Z}{\lambda_{De}} \frac{k \lambda_{De}}{(1 + k^2 \lambda_{De}^2)^{3/2}} . 
\end{equation}

Returning to evaluating equation~(\ref{eq:midqie}), we use spherical polar coordinates for $\vc{k}$, and take the parallel direction along $\vc{v}-\vc{v}^\prime$, so that $\vc{k} \cdot ( \vc{v} - \vc{v}^\prime) = k_\parallel (v - v^\prime) \approx k_\parallel \bar{v}$. Evaluating the $k_\parallel$ integral, the $k^2$ terms become $k^2 = k_\perp^2$. After the azimuthal integral, the only nonvanishing components of the tensor $\Q_{\IE}^{s-s}$ are the $\hat{x} \hat{x}$ and $\hat{y} \hat{y}$ components. Both components have the same magnitude, which we take as the scalar $Q_{\IE}^{s-s}$.  We also apply the variable substitution $\kappa = k_\perp \lambda_{De}$. With these, our scalar estimate for the collisional kernel becomes 
\begin{equation}
Q_{\IE}^{s-s} \approx \frac{\pi q_s^4}{m_s \bar{v}} \int_{\kappa_c}^\infty d\kappa \frac{\kappa^3}{(1 + \kappa^2)^3} \exp \biggl[ \frac{Z}{\lambda_{De}} \sqrt{\frac{\pi m_e n_i}{2 M_i n_e}}  \frac{\kappa}{(1 + \kappa^2)^{3/2}} \biggr] , \label{eq:amidqie}
\end{equation}
in which we have set the lower limit of integration to $\kappa_c$ so that only the unstable $k$ are integrated over. The limit $\kappa_c$ can be determined from the instability criterion $V_i - c_s/\sqrt{1 + \kappa^2} > 0$, which gives 
\begin{eqnarray}
\kappa_c \equiv  \label{eq:kappac}
\left\lbrace \begin{array}{cc}
\sqrt{c_s^2 / V_i^2 - 1}\ , & \textrm{for}\ V_i \leq c_s \\
0\ , & \textrm{for}\ V_i \geq c_s
\end{array} \right.  .
\end{eqnarray}

The $\kappa$ integral in equation~(\ref{eq:amidqie}) will be approximated as follows. The integrand is peaked about the maximum of $\kappa^3/(1+\kappa^2)^3$, which occurs at $\kappa = 1$. Expanding the argument of the exponential about this point yields 
\begin{equation}
\frac{\kappa}{(1+\kappa^2)^{3/2}} \biggl|_{\kappa=1} = \frac{\sqrt{2}}{4} - \frac{\sqrt{2}}{8} (\kappa -1) - \frac{3 \sqrt{2}}{32} (\kappa -1)^2 + \ldots 
\end{equation}
Keeping only the lowest order term of this series, we use the approximation $\kappa/ (1+\kappa^2)^{3/2} \approx \sqrt{2}/4$ in the exponential. The integrand is then algebraic, and can be evaluated analytically
\begin{equation}
\int_{\kappa_c}^\infty d\kappa \frac{\kappa^3}{(1+\kappa^2)^3} = \frac{1}{4} \frac{1 + 2 \kappa_c^2}{(1+\kappa_c^2)^2}  . 
\end{equation}
With this approximation to the $\kappa$ integral, equation~(\ref{eq:amidqie}) becomes
\begin{equation}
Q_{\IE}^{s-s} \approx \frac{\pi q_s^4}{4 m_s \bar{v}} \frac{1 + 2 \kappa_c^2}{(1 + \kappa_c^2)^2} \exp \biggl( \sqrt{\frac{\pi m_e n_i}{16 M_i n_e}} \frac{Z}{\lambda_{De}} \biggr) . \label{eq:qiefinal}
\end{equation}


\end{document}